\newcommand{\pa}{\partial}
\newcommand{\be}{\begin{equation}}
\newcommand{\ee}{\end{equation}}
\newcommand{\bea}{\begin{eqnarray}}
\newcommand{\eea}{\end{eqnarray}}
\def \ci{\cite}
\def \YY {{\rm Y}}
\def \lra {\leftrightarrow}
\newcommand{\nn}{\nonumber}
\newcommand{\p}[1]{(\ref{#1})}
\newcommand{\bt}[1]{{\bar t}}

\def \bran {\bra{n}}
\def \ketn {\ket{n}}

\def \inti { \int^{2\pi}_0 {d\s \ov 2\pi} }

\def \sm {{sigma-model}}

\newcommand{\sn}{\mathop{\mathrm{sn}}\nolimits}
\newcommand{\cn}{\mathop{\mathrm{cn}}\nolimits}
\newcommand{\dn}{\mathop{\mathrm{dn}}\nolimits}

\documentclass[12pt]{article}
\input epsf
\usepackage{epsfig,color,latexsym}
\def \sql {{\sqrt{\l}}\ }
\def \del{\partial}
\def \a {\alpha}
\def \aa {{\a'}}
\def\g{\gamma}
\def\s{\sigma}
\def\z{\zeta}
\def\zi{\zeta_1}
\def\zii{\zeta_2}
\def\ov{\over}
\def\la{\label}
\def\I{{\cal I}}
\def\J{{\cal J}}

\def \om {\omega}
 \def \cH {{\cal H}}
\def\E{{\cal E}}
\def\w{\omega}
\def\b{\beta}
\def\l{\lambda}
\def\eps{\epsilon}
\def\vep{\varepsilon}
\def \adss{$AdS_5 \times S^5$\ }

\def \r { \rho}
\def \sql {\sqrt{\lambda} }
\def \t {\theta}
\def \p {\phi}
\def \vp {\varphi}
\def \Om {\Omega}
\def \ads {{$AdS_5$}}
\def \ov {\over}
\def \s{\sigma}
\def \pa{\partial}
\def \ta{\tau}
\def \sh {\sinh}
\def \ha {{1 \over 2}}

\def \la{\label}

\def \k {\kappa}
\def\foot{\footnote}

\def \const {{\rm const}}

\def \J {{\cal J}}
\def \L {\Lambda}
\def\rr {{\rm r}}

\def \sa {\sum_{a=1}^2}

\newcommand{\rf}[1]{(\ref{#1})}
\renewcommand{\theequation}{\thesection.\arabic{equation}}
\renewcommand{\thefootnote}{\fnsymbol{footnote}}

\def\appendix#1{
  \addtocounter{section}{1}
  \setcounter{equation}{0}
  \renewcommand{\thesection}{\Alph{section}}
  \section*{Appendix \thesection\protect\indent \parbox[t]{11.15cm}
  {#1} }
  \addcontentsline{toc}{section}{Appendix \thesection\ \ \ #1}
  }

\def \four{{\textstyle {1\ov 4}}}
 \def \third { \textstyle {1\ov 3}}
\def\det{\hbox{det}}
\def\be{\begin{equation}}
\def\ee{\end{equation}}

\def \ci {\cite}

\def \foot {\footnote}
\def \bi{\bibitem}
\def \tr {{\rm tr}}
\def \ha {{1 \over 2}}
\def \td {\tilde}

\def \ci{\cite}
\def \N {{\cal N}}
\def \ww {\Omega}
\def \const {{\rm const}}

\def \ss {\sum_{i=1}^3 }

\def \t {\tau}
\def\S{{\cal S} }

\def \nn {\nu}
\def \XX {{\rm X}}
\def \Om {\Omega}
\def \vom {{\bar \omega}}
\def \Y{{\rm Y}}
\def \zz {{\rm z}}
\def \rL {{L}}
\def \ab {{\rm a}}
\def \n {\nu}

\def \rS {{\rm S}}

\newcommand{\bra}[1]{\mbox{$\langle #1 |$}}
\newcommand{\ket}[1]{\mbox{$| #1 \rangle$}}

\newcommand{\ie}{{\it i.e.}}

\newcommand{\vn}{\ensuremath{\vec{n}}}
\newcommand{\tl}{\ensuremath{\tilde{\lambda}}}
\newcommand{\ct}{\ensuremath{\cos\theta}}
\newcommand{\st}{\ensuremath{\sin\theta}}
\newcommand{\ps}{\ensuremath{\partial_1igma}}

\newcommand{\tS}{\ensuremath{\tilde{S}}}

\newcommand{\ctq}{\ensuremath{\cos\theta'}}
\newcommand{\stq}{\ensuremath{\sin\theta'}}
\newcommand{\cdp}{\ensuremath{\cos\Delta\phi}}
\newcommand{\sdp}{\ensuremath{\sin\Delta\phi}}

\newcommand{\beq}{\begin{equation}}
\newcommand{\eeq}{\end{equation}}

\newcommand{\beqa}{\begin{eqnarray}}
\newcommand{\eeqa}{\end{eqnarray}}

\newcommand{\comment}[1]{}
\newcommand{\commentout}[1]{}

\newcommand{\istp}{\ensuremath{\int_0^{2\pi}\frac{d\sigma}{2\pi}}}

\newcommand{\tpJq}{\ensuremath{\left(\frac{2\pi q}{J}\right)}}

\newcommand{\mstrut}[1]{\rule{0pt}{#1 \baselineskip}}

\def \rE {{\rm E}} 
\def \rK {{\rm K}} 

\newcommand{\idop}{\mathbf{1}}

\begin{document}


\vskip-1pt
\hfill {\tt hep-th/0403120}
\vskip-1pt
\hfill BRX TH-537
\vskip0.2truecm
\begin{center}
\vskip 0.2truecm {\Large\bf
Large spin limit of $AdS_5\times S^5$ string theory and \\
\vskip 0.2truecm
low energy expansion 
of ferromagnetic spin chains \\ 
}
\vskip 1.6truecm
{\bf M. Kruczenski$^{1,}$\footnote{E-mail: martink@brandeis.edu},
A.V. Ryzhov$^{1,}$\footnote{E-mail: ryzhovav@brandeis.edu}
and A.A. Tseytlin$^{2,}$\footnote{Also at Imperial College London
and  Lebedev  Institute, Moscow}\\
\vskip 0.4truecm
$^{1}$
{\it  Department of Physics,  Brandeis University\\
Waltham, MA 02454, USA}\\
\vskip .2truecm
$^{2}$ {\it Department of Physics,
The Ohio State University\\
Columbus, OH 43210-1106, USA}\\
 }
\end{center}
\vskip 0.5truecm
\vskip 0.2truecm \noindent\centerline{\bf Abstract}
\vskip .2truecm
By  considering   $AdS_5\times S^5$  string states with large
angular momenta in $S^5$  one is able to provide  non-trivial 
quantitative  checks of the AdS/CFT duality.
A  string  rotating in $S^5$
with two  angular momenta $J_1$,$J_2$  is dual to 
an  operator in $\mathcal{N}=4$ SYM theory whose 
conformal dimension can be computed by diagonalizing 
a (generalization of) spin $1/2$ Heisenberg chain
Hamiltonian.
 It was recently argued
and verified to  lowest order
in a large $J=J_1+J_2$ expansion, that the Heisenberg chain can be
described using a non-relativistic 
low energy effective 2-d 
action for a unit vector field $n_i$ 
which exactly matches the corresponding large $J$ limit of
the classical $AdS_5\times S^5$ string action. 
In this paper 
we show that this agreement extends to the next order and develop a
systematic procedure to compute
higher orders in such large angular momentum expansion. 
This involves several non-trivial steps.
On the string side,  we need to choose a special  gauge 
with a non-diagonal world-sheet
metric which insures  that the angular momentum 
is uniformly distributed along the string, as indeed 
 is the case on the spin chain side.
 We need  also  to 
implement an order by order    redefinition of the field 
$n_i$  to get an action linear in the time derivative. 
On  the spin chain side, it turns out to be crucial to include the
effects of integrating out short 
wave-length modes. 
In this way we gain a better understanding of how
(a subsector of) the string sigma model  
emerges from the dual gauge  theory, allowing us to 
demonstrate the duality beyond comparing 
particular examples of states with large $J$.

\newpage

\renewcommand{\thefootnote}{\arabic{footnote}}
\setcounter{footnote}{0}

\vskip 0.5truecm
\vskip 0.5truecm
\vskip 0.5truecm

\def \Y{{\rm Y}}

\def\[{\begin{equation}}
\def\]{\end{equation}}
\def\<{\begin{myeqnarray}}
\def\>{\end{myeqnarray}}

\def \lc {light-cone\ }
\def \sm {$\s$-model }
\def \la {\langle}
\def \ra {\rangle}
\def \e {{\rm e}} \def \four{{\textstyle{1\ov 4}}}
\def \ov {\over}
\def \we { \wedge}
\def \F {{\mathcal F}}
\def \ep {\epsilon}
\def \k {\kappa}
\def \N {{\mathcal N}}
\def \L {{\mathcal L}}
\def \K {{\mathcal K}}
\def \I {{\mathcal I}}
\def \J {{\mathcal J}}

\def \a {\alpha}
\def \E {{\mathcal E}}
\def \b {\beta}
\def \g {\gamma}
\def \G {\Gamma}
\def \d {\delta}
\def \l {\lambda}
\def \La {\Lambda}
\def \m {\mu}
\def \n {\nu}
\def \s {\sigma}
\def \S {\Sigma}
\def \r {\rho}
\def \t {\theta}
\def \ta {\tau}
\def \p {\phi}
\def \P { \Phi}
\def \vp {\varphi}
\def \ev {\varepsilon}

\def \rL   {{\rm L}}
\def \frac#1#2{{ #1 \over #2}}
\def \td {\tilde}
\def \M {{\mathcal M}}
\def \aa {{\a'}}

\def \adss {$AdS_5 \times S^5\ $}
\def \ads {$AdS_5$\ }
\def \pw {plane wave\ }
\def \N {{\mathcal N}}
\def \lc {light-cone\ }
\def \ta { \tau}
\def \s { \sigma }
\def  \sqf {\l^{1/4}}
\def \sg {\sqrt {g }}
\def \vp {\varphi}
\def \fourth {{1 \ov 4}}
\def \fo  {{{\textstyle {1 \ov 4}}}}
\def \inv {^{-1}}
 \def \diag {{\rm diag}} \def \td { \tilde }

\def \ab {{\rm a}}

\def \la {\label}
\def \alpr {\alpha'}
\def \om {\omega}
\def \del{\partial}
\def \Tr {{\rm Tr}}
\def \R {R^{(2)} }
\def \la {\label}
\def \tr {{\rm tr}}
\def \ha {{1 \over 2}}
\def \ov {\over}
\def \JJ {{\mathcal J}}
\def \Om {\tilde \Omega}
\def \ome {\Omega}

\def \z {\zeta}
\def \w  {\omega}
\def \oo  {\omega}
\def \ww { {\rm w} }
\def \www { {\rm w}_{21} }
\def \Ev {E_{\rm vac}}
\def \sql {{\sqrt{\l}}\ }
\def \tri {{\textstyle {1\ov 3}}}
\def \ta {\tau}
\def \tdr {{\td \r}}
\def \ap {\approx}
\def \isql {{\textstyle { 1 \ov \sqrt{\l}}}}

\def \D {{\Delta}}
\def \ta {\tau}

\def \sp {$S^5$ }

\def \xs{\xi}
\def \kk {{\rm k}}
\def \po {{\psi_0}}
\def \sql {{\sqrt{\l}}\ }
\def \del{\partial}

\def\E{{\mathcal E}}
\def\w{\omega}
\def\b{\beta}
\def\l{\lambda}
\def\eps{\epsilon}
\def\vep{\varepsilon}
\def \rL {{\rm L}}

\def \ads {{$AdS_5$}}
\def \ov {\over}
\def \s{\sigma}
\def \pa{\partial}
\def \ta{\tau}
\def \sh {\sinh}
\def \ha {{1 \over 2}}

\def \k {\kappa}
\def \const {{\rm const}}

\def \four{{\textstyle {1\ov 4}}}
 \def \third { \textstyle {1\ov 3}}
\def\det{\hbox{det}}
\def\be{\begin{equation}}
\def\ee{\end{equation}}

\def \foot {\footnote}
\def \bi{\bibitem}
\def \tr {{\rm tr}}
\def \ha {{1 \over 2}}
\def \td {\tilde}
\def \ci{\cite}
\def \N {{\mathcal N}}
\def \ww {\Omega}
\def \const {{\rm const}}
\def \ss {\sum_{i=1}^3 }
\def \t {\tau}
\def\S{{\mathcal S} }
\def \nn {\nu}
\def \XX {{\rm X}}

\newcommand{\nln}{\nonumber\\}
\newcommand{\nl}{\nonumber\\&&\mathord{}}
\newcommand{\nlnum}{\\&&\mathord{}}
\newcommand{\nle}{\nonumber\\&=&\mathrel{}}
\newcommand{\earel}[1]{\mathrel{}&#1&\mathrel{}}
\newenvironment{myeqnarray}{\arraycolsep0pt\begin{eqnarray}}{\end{eqnarray}\ignorespacesafterend}
\newenvironment{myeqnarray*}{\arraycolsep0pt\begin{eqnarray*}}{\end{eqnarray*}\ignorespacesafterend}

\newcommand{\lrbrk}[1]{\left(#1\right)}
\newcommand{\bigbrk}[1]{\bigl(#1\bigr)}
\newcommand{\brk}[1]{(#1)}
\newcommand{\vev}[1]{\langle#1\rangle}
\newcommand{\normord}[1]{\mathopen{:}#1\mathclose{:}}
\newcommand{\lrvev}[1]{\left\langle#1\right\rangle}
\newcommand{\bigvev}[1]{\bigl\langle#1\bigr\rangle}
\newcommand{\bigcomm}[2]{\big[#1,#2\big]}
\newcommand{\comm}[2]{[#1,#2]}
\newcommand{\lrcomm}[2]{\left[#1,#2\right]}
\newcommand{\acomm}[2]{\{#1,#2\}}
\newcommand{\bigacomm}[2]{\big\{#1,#2\big\}}
\newcommand{\gcomm}[2]{[#1,#2\}}
\newcommand{\lrabs}[1]{\left|#1\right|}
\newcommand{\abs}[1]{|#1|}
\newcommand{\bigabs}[1]{\bigl|#1\bigr|}
\newcommand{\bigeval}[1]{#1\big|}
\newcommand{\eval}[1]{#1|}
\newcommand{\lreval}[1]{\left.#1\right|}

\def \lra {\leftrightarrow}
\def \vom {{\bar \omega}}
\def \E {{\mathcal  E}} \def \J {{\mathcal  J}}
\def \YY {{\rm Y}}

\def \d {\del}
\def \rJ {{J}}
\def \Q {{\cal S}}
\def \sms {sigma models\ }
\def \sm {sigma model\ }
\def \L {\Lambda}
\def \gl {\ell}
\def \tr {{\rm tr\ }}
\def\z{\zeta}
\def\zi{\zeta_1}
\def\zii{\zeta_2}
\def\K{\mbox{K}}
\def\eE{\mbox{E}}   \def \vt {\vartheta}
\def \vr {\varrho}
\def \wup {w}

\renewcommand{\theequation}{\thesection.\arabic{equation}}
\renewcommand{\thefootnote}{\fnsymbol{footnote}}

\def \sn {{\rm sn}}
\def \dn {{\rm dn}}\def \cn {{\rm cn}}

\def \vo{\omega}
\def \wup {w}
\def \half {\ha}
\def \cN {{\mathcal N}} \def \rN {{\rm N}}
\def \rr {\vr}

\def \rr{{\rm r}}
\def \zz{{\rm z}}
\def \q{{\rm q}}
\def \sa {\sum_{a=1}^2}
\def \lc {light-cone\ }
\def \ta{{\rm t}}
\def \fo {{1 \ov 4}}
\def \rJ  {{J}}
\def \J {{\cal J}}
\def \bl {{\tilde \l}}
\def \D {{\cal D}}
\def \four { {1 \ov 4}}
\def \HH {{\cal H}}
\def \rH {{\rm H}}
\def\d {\del}
\def \ss{{\rm s}}
\def \te {{\rm t}}

\def \rS {{\rm S}}
\def \tl{{\bl}}


\setcounter{equation}{0}
\setcounter{footnote}{0}

\setcounter{equation}{0}
\setcounter{footnote}{0}
\section{Introduction  } \label{intro}

Understanding AdS/CFT duality beyond the BPS or near BPS \ci{bmn}
limit remains an important challenge.
It was suggested in \ci{gkp} that concentrating on string states
with large quantum numbers, like angular momentum in $AdS_5$,
one finds   a qualitative (modulo interpolating function of
`t Hooft coupling $\l$)  agreement between the
 $AdS_5$ string energies and anomalous dimensions of the corresponding
gauge theory operators (see also \ci{ft1,tse1,kru2}).
About a year ago,  it was observed \ci{ft2} that semiclassical string
states  with several non-zero angular momenta (with large total $S^5$
momentum $J$)  have a remarkable property that their energy
admits an  analytic expansion in $\bl\equiv {\l\ov J^2}$ at large $J$.
\foot{Earlier examples of
similar  solutions were found in \ci{ft1,rus,min}.}
It was proposed, therefore, that the coefficients 
of such an expansion can be matched precisely  with the perturbative anomalous
dimensions of the corresponding scalar
SYM  operators computed in the same  $ J \to \infty,
\bl < 1$ limit \ci{ft2}. That  would provide the first quantitative check of
AdS/CFT duality far from the BPS limit.
The reason  for this expectation   was that
for such special solutions all  string $\a'\sim {1 \ov \sql}$ 
corrections might   be suppressed
in the large $J$ limit (as was explicitly checked
for a particular case in \ci{ft3}; see also \ci{tse2} for a review). Then, the classical string energy would
represent an exact string theory  prediction in this limit.
This proposal received a spectacular confirmation in \ci{bmsz,bfst}
where the  one-loop anomalous dimensions of  the relevant
scalar SYM operators were computed utilizing a  remarkable
  Heisenberg spin chain interpretation of the one-loop anomalous dimension
 in the scalar sector  \ci{mz1} and taking the  thermodynamic $J \to
 \infty$  limit of the Bethe ansatz solution for the eigenvalues.
The detailed agreement of the functional dependence of the 
leading ``one-loop''
coefficient  on the ratio of spins for ``inhomogeneous'' folded and
circular rotating string solutions
was further demonstrated in
\ci{ft4,afrt,bfst}. Same was  found 
 also for  the  ``homogeneous'' \ci{art}
two-spin circular solutions  \ci{kmmz} and for  particular three-spin
states  \ci{emz,kri}.
This agreement was extended to the ``two-loop''
level using integrable model/Bethe ansatz techniques \ci{serb,kmmz}.

One would  obviously like to achieve a
better understanding of    how and why
this correspondence between the  string theory and gauge theory
 works, e.g.,   the general rules of
how particular    string states are mapped onto particular SYM states.
 It would be interesting
 to see how string \sm world-sheet action emerges
on the gauge theory side (cf. \ci{bmn}), allowing one to go
beyond discussion of matching of individual  states.
An important suggestion in this direction  was made recently in \ci{kru},
and our  aim here will be to further
clarify  and  extend it  beyond the
leading  (``one-loop'') order.

In more detail, 
  here we  would like to  try to  understand  in
general the  correspondence
between (``semiclassical'')
string states  with two large 
angular momenta $J_1,J_2$  in $S^5$ 
 and  single-trace SYM operators
 $O_{J_1,J_2}=\tr( \Phi_1^{J_1} \Phi_2^{J_2} +\ldots)$
 ($\Phi_1,\Phi_2$ are two complex combinations
of SYM scalars).
The main assumption  is that the limit
\be \la{lim}
J\equiv J_1 + J_2 \gg 1 \ , \ \ \ \ \ \ \
\bl \equiv {\l \ov J^2} ={\rm fixed}  < 1 \ , \ee
i.e.  the expansion in powers of $1\ov J$ and $\bl$
is well-defined on both the string and the SYM sides of the duality.
The classical energy of such  rotating string solutions
admits  the following expansion \ci{ft2}
\be \la{ne}
E= J\ F({J_2\ov J},\bl) \ , \ \ \ \ \ \ \ \ \ \ \ \ 
 F=  1 + c_1  \bl + c_2 \bl^2 + \ldots\ ,  \ee
where $c_i$ depend on ratios of spins (and other
parameters like winding numbers). If  quantum
string corrections to $c_i$ are suppressed by extra powers
 of $1\ov J$ \ci{ft3}, the classical string energy
  \rf{ne} should
 represent the exact string result in the large $J$ limit.
 Then the AdS/CFT duality  implies
 that one should be able to match \rf{ne}   with
  dimensions of the corresponding SYM operators
 found in the same limit.
 Indeed, it   was  demonstrated in
 \ci{bmsz,ft4,afrt,bfst} and \ci{serb,kmmz}
 that  energies of a particular  classical rotating 2-spin string
solutions agree precisely  with anomalous dimensions
of the corresponding SYM operators at the first two --
``one-loop'' and ``two-loop'' --  orders in expansion in $\bl$
at large $J$. There is also matching at the level of integrable
structures \ci{as,emz} clarified and established  in general 
in \ci{kmmz}.

 The one-loop  anomalous dimension matrix in the
  sector of 2-spin  operators $O_{J_1,J_2}$  happens
to be equivalent to the ferromagnetic  Heisenberg
XXX$_{1/2}$ ($SU(2)$)  spin chain Hamiltonian $\rH$ \ci{mz1}.
With the 2-loop \ci{bks} and  3-loop \ci{bks,bei,beit}
 corrections  included $\rH$   may be interpreted  as a
generalized spin chain Hamiltonian containing further
 next to nearest neighbor  interactions.
To find the  eigenvalues of $\rH$ in  the one-loop approximation
one is able to apply the   Bethe  ansatz techniques
 with  crucial simplification of thermodynamic limit $J \to \infty$
\ci{bmsz,bfst,kmmz}. Furthermore,
one is able  to extend this  to the two-loop
level by embedding \ci{serb} the anomalous
 dimension operator into a
particular integrable spin chain system
 and then again utilize 
  the thermodynamic limit of the Bethe ansatz
 \ci{serb,kmmz}.
 Using the Bethe ansatz for the spin chain 
 on the gauge side and integrability 
 of the classical sigma model on the string side  ref.\ci{kmmz} 
 managed to prove the one-loop and two-loop matching  for generic
 solutions. 
 
Our approach below will be  simpler and in a sense complementary to the 
one of \ci{kmmz} based on integrability.
Following 
\ci{kru}, we   would like to establish a  more direct way
of comparing the string and SYM results without the  need to 
go  through detailed 
analysis  of particular   solutions (or their general classification)
and 
 use of complicated Bethe ansatz techniques which seem
 hard to extend beyond few leading orders without knowledge of
 integrability of  the exact spin chain Hamiltonian.
 The idea is to
extract an  effective  
2-d  action  describing
low-energy states of the ferromagnetic Heisenberg
spin  chain in the limit $J \to \infty $, \ $\bl =$fixed
and then  to show that it agrees with the  string \sm action expanded
in the same limit.
On the string side that would
correspond to ``gauging away'' a collective
coordinate associated with the total (orbital and internal) $S^5$ spin
$J$ (which   may be viewed as  a
 non-trivial generalization
 of \lc gauge fixing in the BMN case \ci{bmn}).\foot{This amounts
 to a reorganization of the classical
string action  in the  large $J$ sector so that the  expansion effectively
goes in powers of  $\bl\sim \l $  and thus
corresponds  to a  ``near zero-tension'' \ci{matt}  limit.
Related large $J$ limit in the classical string equations
was considered from a geometric point of view in \ci{mik}.}
An agreement between the resulting effective spin chain and 
string
actions would  then  imply an agreement
between energies of  particular solutions/states
\comment{NEW}
(as well as matching of underlying integrable structures).
As a consequence, one can
directly  relate the target space spinning string
 configurations to
configurations  of 1-d spins \ci{kru},
 a  non-trivial connection which emerged 
   also in the Bethe ansatz approach 
  \ci{bmsz,kmmz}.
\comment{NEW} Once the configurations are related, many questions, as
for example the agreement between Bethe-Ansatz and sigma model calculations,
between integrable structures etc. become questions regarding the 
spin chain Hamiltonian and not the AdS/CFT correspondence. By that we mean that
the agreement is between two different ways of describing a spin chain, a
low energy description in terms of an effective action or an exact description
using the integrability properties of the model.
 The AdS/CFT correspondence simply establishes that one of those two ways
is directly related to (a limit of) the action of a string moving in a specific 
target space.  

By starting with the one-loop expression for the dilatation operator 
  \ci{mz1}
\be \la{hei}
\rH=  {\l \ov (4 \pi)^2}
 \sum^J_{a=1} ( 1- \s_a \cdot \s_{a+1}) \ ,
 \ \ \ \ \ \ \ \  \s^i_a \s^j_a  = \delta^{ij} + i \ep^{ijk} \s^k_a 
  \ ,   \ee
 one finds \ci{kru} (see, e.g., \ci{fra,ran} and refs. there)
 that the corresponding action in
 coherent state  \ci{pere}
 ($\bra{n} \s^i_a\ket{n} = n^i_a$, \  $(n^i)^2=1, $ $i=1,2,3$)
 path integral is
 \be \la{kh}
 \rS= \int dt\  \bigg[
   \sum^J_{a=1} L_{\rm WZ} (n_a) - \bra{n}\rH\ket{n} \bigg] \ ,\ee
   \be \la{keh}
 \bra{n}\rH\ket{n} = {\l \ov 2(4 \pi)^2} \sum^J_{a=1} (n_{a+1}
 - n_a)^2  \ ,      \ee
 where $L_{\rm WZ} (n_a)$ depends on $n_a$ only at a given site
  $a$ and is  linear  in $\d_t n_a$.\foot{$L_{\rm WZ}$  may be 
  viewed as an
  analog of the usual ``$p \dot q$'' term,
  i.e. this non-relativistic action may be interpreted as
  a phase space action with $n_i$ describing both coordinates and
   momenta. This ensures proper commutation relations if  
 we reverse  the logic and promote  $n_i$ to quantum operators:
 $n_i \to \hat n_i \equiv \s_i, \ [\hat n_i,\hat n_j] =
  2 i \ep_{ijk} \hat n_k$.}
  Since we are interested in the limit of $J \to \infty$
with $\bl$ fixed  this  suggests to take the continuum limit,
introducing $n(\s)$,
 $ 0 < \s \leq  2\pi, $  with $ n_a= n({2\pi a \ov J} )$.
We then  finish with
\be \la{gh}
 \rS=  J  \int dt \inti \ L \ , \ \ \ \ \ \ \ \
 L= C_t (n)- {1 \ov 8}  {\bl} ( \del_1 n)^2  \ , \ \ \ \ \ \ \  \del_1 \equiv
 \del_\s \ ,  \ee
 where we set
$L_{\rm WZ} (n)  \equiv  C_t (n) $  to indicate that this term 
 is linear in $ \d_t n^i.$
 The  equation of motion
corresponding to   \rf{gh}
are the classical ferromagnet or Landau-Lifshitz (LL)  equation\foot{The
LL equation describes evolution in time
of magnetization vector of a (one-dimensional  in the present context)
macroscopic ferromagnet (see, e.g., \ci{fra,kos}).}
\be \la{ll}
\d_t n_i = \ha \bl\ \ep_{ijk} n_j \d^2_1 n_k \ . \ee
 We have omitted higher-derivative terms  coming from
  $(n_{a+1} - n_a)^2$ since they are  suppressed by powers of $1\ov J$.
  We further observe that since $J$ is the coefficient 
   in front of
  the action, it   plays the role of the inverse
  Planck constant in the spin chain path integral.
  Then  in the limit  when
  $J\to \infty$ and $\bl=$fixed one should be  able to ignore quantum
  corrections  and thus to treat \rf{gh}  as a classical action that
  should be matched to   the  string action
  expanded in the same limit. This matching was indeed
  demonstrated  in  \ci{kru} and will be further discussed
  and extended to higher  orders in $\bl$  below.

\commentout{

With   higher-order in the `t Hooft coupling 
$\l$ (``higher-loop'')  corrections 
to the large $N$  SYM dilatation operator included in \rf{hei}, 
it  takes  again the form of a spin chain 
Hamiltonian \ci{bks,beit}  containing at order $\l^n$ 
 up to $(n+1)$-neighbor interactions  between  quantum spins
 at $J$ sites. 
We expect, therefore, that  there should  exist  a semiclassical 
 low energy effective theory description for 
this spin chain as $J \to \infty$.
We may  then attept to  calculate a Wilsonian low energy effective 
action (or effective Hamiltonian) governing the slow spin wave modes,
 by starting with the  full quantum 
partition function and integrating out the fast modes. 
 Starting with spin chain Hamiltonian $\rH$ containing a single sum over
spins at adjacent sites, we  should 
 then expect to find an effective  Hamiltonian 
$H$ given by  a single integral over spatial direction and depending on  
 unit vector field  $n$ and its spatial derivatives.
 The time derivative term in the effective action should be still 
 the WZ term $C_t(n)$ (its coefficeint is fixed by 
 the semiclassical spin quantization conditions or topological argument).  
 
}

We should remark that at higher loops, the operators of interest are still described by
a spin $s=\half$ chain and only the Hamiltonian gets modified. At each loop
interactions involving larger number of neighbors are introduced. However, at
least at the next order, the interaction is still ferromagnetic (for small $\bl$)
and therefore, at low energy, the description of the system in terms of long-wave 
length spin waves should be valid. The low energy effective action 
governing these modes contains, a priori, all possible terms compatible with the symmetries. 
Our task is then to compute the corresponding coefficients which  
should include the effects of integrating out the high momentum modes.

Thus, with  higher-loop   corrections 
included in $\rH$ in \rf{hei}, we  expect 
to find a low-energy effective action  analog of \rf{gh},\rf{ll}
with higher powers of $\bl$ 
multiplying higher derivatives of $n_i$ 
\be\la{expe}
L= C_t (n) - H  (\d_1 n, \d^2_1 n, \d^3_1 n,\ldots) \ , \ee
$$
H
= a_0 \bl  ( \del_1 n)^2
+ \bl^2 \bigg[ a_1 ( \del^2_1 n)^2 + a_2 ( \del_1 n)^4\bigg]
$$  \be\la{xpe} + \
 \bl^3 \bigg[ a_3 ( \del^3_1 n)^2 + a_4 ( \del_1 n)^2 ( \del^2_1 n)^2
 + a_5 ( \del_1 n \del^2_1 n)^2  + a_6 ( \del_1 n)^6 \bigg] +  O(\bl^3)  
  \ . \ee
 We have written down all possible local structures
 with 4 and 6 spatial derivatives  built out of a unit vector
  field $n_i(t,\s)$  (modulo integration by parts).
The coefficients $a_0,a_1,a_2, ...$ are determined by  detailed 
microscopics of the spin chain.\foot{In principle, one could 
use the knowledge 
of a few families of microscopic solutions to fix these 
coefficients, but it turns out one can work with general field 
configurations, as we will show in Section \ref{sec:qc}.}

 It is useful to note  that since   \rf{ll} and $L$ in  \rf{gh}
 or \rf{expe}
are linear in time derivatives, the  overall factor of
$\bl$ in $H$ can be
absorbed into a rescaling of the time coordinate
 (this  does not change the coefficient in front of the action
 \rf{gh})
 \be \la{resca}  t \to \bl^{-1} t \ , \ \ \ \ \
 L \to L = C_t (n) - 
a_0 ( \del_1 n)^2
-  \bl \big[ a_1 ( \del^2_1 n)^2 + a_2 ( \del_1 n)^4\big]
+  \ldots \ . \ee
Our first aim below will be to show, generalizing the 
suggestion of  \ci{kru},
 how  such an action with a single time derivative  (in $C_t (n)$ term) 
 but containing all orders of
derivatives
in $\s$,   emerges
 from the usual second-time-derivative  \adss \sm
 action expanded  in the limit $J \to \infty, \bl < 1$.
 In the   2-spin sector the classical
string equations in \adss expanded for large spin
reduce to a higher-derivative generalization
of the LL  equation for a unit
 3-vector $n_i$ describing a shape of a string 
 rotating  in two orthogonal planes.
It is   remarkable that  such a
 non-relativistic 
 ``classical  ferromagnetic''
 action containing all orders in spatial derivatives
 but only first order in time derivative happens to be
   a particular ``re-expansion'' (in the $\bl \to 0$ limit)
 of the usual relativistic string sigma model action.

 Next, we will  compare the coefficients in $H$ \rf{xpe}
 appearing from  the (quantum) spin chain 
  with the corresponding
 coefficients coming out of the string \sm  action.  
 We shall conclude, 
   in full  agreement with the previous results based on Bethe
 ansatz technique \ci{serb,kmmz}, that 
 the correspondence does extend to  the next  $\bl^2$
 (``two-loop'')  order.
  We shall find that beyond the leading $\bl$ order one
 is not able to ignore quantum corrections 
 on the spin chain side:
  before taking  the continuum limit one should 
 compute a quantum
 effective action analog of \rf{kh}. 
 While the terms quadratic
 in $n_i$ in \rf{xpe} are indeed correctly  found    by
 simply taking the continuum limit of  the  coherent state
  expectation value of the spin chain Hamiltonian 
  equal to the SYM  dilatation operator,
  to reproduce the $n^4$ and higher order terms one  needs
  to take into account quantum corrections. 
  We shall also discuss  how the matching
 should work at the ``three-loop'' $\bl^3$ order  but
 detailed quantum computations on the spin chain side 
 (beyond the evaluation of the corresponding coherent 
 state expectation
 value $ \bra{n}\rH\ket{n}$)  will not   be described  here.


 The rest of this   paper is organized as follows. 
In  section \ref{stringaction}  we discuss  the large angular momentum
 limit of the string action. We first  consider the  conformal 
gauge (section \ref{sec:confgauge})  and obtain the  expansion 
of the action in powers of $\bl$. 
Although this approach  happens to be enough 
up to second order in $\bl$,   at
 higher orders the action   contains  non-local terms. For 
 that reason in section \ref{sec:nondiaggauge}  we  develop a 
more systematic expansion by finding
an appropriate ``uniform'' gauge and applying  an order by order
field  redefinition to eliminate terms which are non-linear 
 in   time derivatives. 
 In section \ref{sec:expe} we  
 take the continuum limit of the coherent 
 state expectation value $\bra{n}\rH\ket{n}$ of the 
spin chain Hamiltonian representing the SYM dilatation 
operator with two-loop and three-loop 
corrections included.
We show, however, that this naive approach 
 does not reproduce the full string results at $\bl^2$ and $\bl^3$ orders.
  In section \ref{sec:qc} we compute quantum corrections to the
 effective action of the spin chain coming from
integrating out high energy modes. 
They turn out to contribute starting with 
  $\bl^2$  order  and 
after including them we find perfect agreement with the string side
at this order. 
 In section \ref{folded} 
 we perform some checks of the $\bl^2$ result 
 for particular two-spin string configurations. 
 Concluding remarks are made in section \ref{remarks}.
 Appendices contain  some   technical details and 
 useful relations.

\setcounter{equation}{0}
\setcounter{footnote}{0}

\section{Generalized ``classical magnetic''
 action from \\
 string \sm on $R\times S^3$ } \label{stringaction}

 Our first task will be to show how the action \rf{expe}
 appears from the standard string sigma model action on
 $R \times S^3$. Here $R$ factor represents the time direction in
  $AdS_5$ and
 $S^3$ factor is from  $S^5$: we shall consider
 only  string configurations
 that are located in the center of $AdS_5$ and belong to
  $S^3 \subset S^5$.
   They may  carry two out of three
 possible  independent  angular momenta and should describe string
 states corresponding to eigenstates of closed $SU(2)$
 subsector of anomalous dimension matrix  or
 spin chain Hamiltonian. 
 There exists a
   generalization of the procedure described below
to  the 3-spin ($SU(3)$) sector but it will not be discussed here.

 \setcounter{footnote}{0}
 \subsection{Parametrization of $R \times S^3$} \label{sec:param}

 The metric of  $R \times S^3$ space-time  can be parametrized as follows
 \be
  ds^2 = - dt^2 + |d\XX_1 |^2 + |d\XX_2 |^2  \ , \   \ \ \ \ \ \
 |\XX_1 |^2 + |\XX_2 |^2 =1 \ ,  \ee
 \be \la{fre}
\XX_1 \equiv X_1 + i X_2 =  \cos \psi \ e^{ i \vp_1} \ , \ \ \ \
\XX_2 \equiv X_3 + i X_4 = \sin  \psi \ e^{ i \vp_2} \ . \ee
In this parametrization string states that carry
two angular momenta $J_1,J_2$  should be   rotating
in the two orthogonal planes $(X_1 , X_2)$ and $(X_3 , X_4)$
(see \ci{ft2,tse2} for  details).
To consider the limit of large total spin  $J=J_1 + J_2$ we
would like to isolate the corresponding collective coordinate,
i.e. the common phase of $\XX_1$  and $\XX_2$.
In the familiar case of  fast motion of the center of mass the role
of $J$ is played by linear momentum or  $p^+$.
Here, however,  $J$ represents
the  sum of ``orbital''  as well as ``internal''
angular momentum   and thus
does not correspond  simply to the center of mass motion.
This is thus a
generalization of the limit considered in \ci{bmn}:
we are interested in ``large''
extended string configurations and not in a
nearly point-like strings.
Let us thus set
 \be \la{freh}
\XX_1  =  U_1  \ e^{ i \a} \ , \ \ \ \
\XX_2 = U_2  \ e^{ i \a} \ , \ \ \ \ \ \ \ \ \
|U_1|^2 + |U_2|^2 =1 \ ,  \ee
i.e.  parametrize $S^3$ in terms of $CP^1$ coordinates $U_i$
 and an angle $\a$ (Hopf $S^1$ fibration of $S^3$).
 The angle $\a$ representing simultaneous rotation in the two planes
  will be the collective coordinate corresponding
 to
 $J$.
  In terms of standard $S^3$ angles
\be \la{be}
  U_1 =  \cos \psi \ e^{ i \b}\ , \ \ \ \
 U_2 =  \sin \psi \ e^{ - i \b}\ , \ \ \ \ \ \ \a= \ha ( \vp_1 +
 \vp_2)  \ , \ \ \ \  \b= \ha ( \vp_1 -\vp_2)  \ .
  \ee
  Then
\be \la{ppi}
 (ds^2)_{S^3}= (d\a - i  U^*_r dU_r)^2  +
 dU^*_r dU_r +  ( U^*_r dU_r)^2 \ , \ \ \ \ \ \ \  r=1,2\ ,  \ee
 i.e.
$$
 (ds^2)_{S^3}= (D\a)^2  + D U^*_r D U_r \ ,
 $$
  \be \la{pp}
D\a \equiv  d \a + C  \ , \ \ \ \ D U_r = dU_r  - i C  U_r  \ ,\ \ \
C\equiv  - i  U^*_r dU_r \ .  \ee
It is useful to replace  $U_r$ by a unit vector $n_i$
representing $CP^1 $
 \be \la{gyg}
n_i \equiv  U^\dagger \sigma_i U \ , \ \ \ \ \ \ \ \
U=(U_1,U_2)  \ ,
\ee
where  $\sigma_i$ are Pauli matrices. Then
\be
(ds^2)_{S^3}=  (D\a)^2  + \fo   dn_i dn_i
\ , \ \ \ \ \ \ \ \ \ \ \    D\a = d\a + C(n) \ ,  \ee
where
 $C(n)$ has a non-local WZ-type representation
$
 C = -\ha  \int^1_0  d \xi\  \ep_{ijk} n_i  \del_\xi
  n_j  d n_k.$\foot{One can check
this relation
using $\s_i^{ab} \s_i^{cd} = - \delta^{ab} \delta^{cd}
+ 2 \delta^{ad} \delta^{bc}$. $C$ may be interpreted as a vector potential 
of a Dirac monopole at the origin.}
In terms of $S^3$ angles  one has
\be \la{nel}
n_i = (\sin 2 \psi \ \cos 2\b;\   \sin 2 \psi \ \sin 2\b;\
\cos 2\psi) \ ,   \ee
 \be
(ds^2)_{S^3}= (d\a + C)^2  +  d\psi^2 + \sin^2 2\psi\ d\b^2 \ , \ \ \
\ \ \ \ \     C= \cos 2 \psi \ d  \b \ . \ee
 It is interesting to note a direct analogy  between \rf{gyg}
relating $n_i$ and $U_1,U_2$
and the coherent state basis on the spin chain side.\foot{
In the spin chain case we have complex scalars $\Phi_1,\Phi_2$
in the operator
tr$(\Phi_1^{J_1}\Phi_2^{J_2}+ \ldots)$ representing spins up and down.
In coherent state basis $\bra{n}\s^i \ket{n} = n^i$ where total
$\ket{n}$ is a product
of doublet coherent states at each node.
We may thus view $U_r$ as ``radial''
coordinates directly corresponding to the two complex scalars
$\Phi_1,\Phi_2$ on the SYM side, with
$ U^\dagger \sigma^i U = n^i$
being the ``classical'' analog of  $\bra{n}\s^i \ket{n} = n^i$.
The $U(1)$  phase $\a$ corresponds to rotating the
chiral superfields by the same  overall phase.}

The string action is then (we use signature $(-+$))
\be \la{ste}
I= { \sql} \int d\tau \int^{2\pi}_0  {d \s\ov 2\pi}  \ L \ , \ee
\be \la{st}
L =  - \ha \sqrt{ - g} g^{pq}
\big( - \d_p t \d_q t + D_p \a  D_q \a  + \fo \d_p n_i \d_q n_i
\big)  \ , \ee
where
\be \la{cce} D_q \a= \d_q \a + C_q (n) \ , \ \ \ \ \
C_q = - \ha \int^{1}_0 d \xi  \ep_{ijk} n_i \del_\xi n_j \del_q n_k
\ , \ee $$
n_i(\tau,\s, \xi=1) = (n_i)_0\ , \ \ \
n_i(\tau,\s, \xi=0) = n_i (\tau,\s)
\ ,  $$
\be \la{cec}\del_p C_q - \del_q C_p =
\ha \ep_{ijk} n_i \del_p n_j \del_q n_k \ ,\ \ \ \ \ \ \
\delta C_q = \ha \ep_{ijk} \delta n_i n_j \del_q n_k + \del_q \chi
\ . \ee
The crucial point is that 
 one  should view  $t$ and $\a$ as ``longitudinal'' coordinates
that reflect
the redundancy of the reparametrization-invariant
string description: they are not ``seen''
on the gauge theory side, and should be gauged away (or
eliminated using the constraints).
At the same time,  the unit vector   $n_i$ should be interpreted as
 a ``transverse'' or physical
 coordinate which should thus  have a counterpart
 on the $SU(2)$ spin chain side (with an  obvious candidate
 being  a vector parametrizing the coherent state).
 To put  \rf{st} into first-time-derivative form like
 \rf{expe} one will need to  properly expand
   the action
 and make a field redefinition of $n_i$.

 The conserved  charges corresponding to translations
 in time, rotations of  $\a$  and $SO(3)$ rotations of $n_i$
 are\foot{Note that $\sql$ is the effective string tension in
 \rf{ste} and so all classical string charges are proportional to
 $\sql$.}
 \be\la{cha} (E,J,\rS_i)= \sql (\E,\J,\Q_i)\ , \ \  \ \ \ \ \ \
\E= - \inti \sqrt{ - g} g^{0p}\del_p t \ , \ \ \
\ee
\be  \la{kop}\J= - \inti  \sqrt{ - g} g^{0p} D_p \a  \ , \ \ \ \ \ \
\Q^i = - \inti  \sqrt{ - g} g^{0p} q^i_p \ , \ee
where the local current $q^i_p$ is
\be\la{okk}
  q^i_p \equiv
   D_p \a \ n_i + \ha \ep_{ijk} n_j \del_p n_k \ . \ee
Given a generic string configuration, one can  apply a
 global rotation to
put $\rS_i$ in the ``canonical'' form $(0,0,\rS_3)$, where
$\rS_3$ will then correspond
 to the difference $J_1-J_2$ of the two $S^3$ spins
 whose sum is $J$ (the corresponding angular
  coordinate is $\b$ in \rf{be}).
  
Note that the effective coupling constant $\bl$ in \rf{lim} is
directly related to the (rescaled) 
charge  $\J$ in \rf{kop}
\be\la{lama}
\bl \equiv {\l \ov J^2} = { 1 \ov \J^2} \ ,  \ \ \ \ {\rm i.e.} 
\ \ \ \ \ 
\J = { 1 \ov \sqrt{\bl} } \ . \ee
Thus expansion in powers of $1\ov \J^2$ is the same as 
expansion in powers of $\bl$.

 \setcounter{footnote}{0}

 \subsection{Conformal gauge choice} \label{sec:confgauge}

While the final expressions for the physical quantities
for specific string solutions like
the  energy as a function of spins
should not of course depend  on a
particular choice of reparametrization gauge, the simplicity of the
``off-shell''
correspondence between the string and spin chain 2-d actions
is  sensitive to  a choice of world-sheet coordinates.
 It turns out that  the simplest conformal gauge choice
fails to be the adequate one beyond the leading order
in $\bl$  expansion: one needs  to choose instead  a non-trivial
(``non-diagonal'')  gauge.
 This  may be viewed as a technical complication,
but it actually
highlights the importance of a suitable  reparametrization gauge
choice in understanding how string action originates
from gauge theory.

Nevertheless, it is still  instructive to start with the discussion
of the standard conformal gauge
$\sqrt{ - g} g^{ab} = \eta^{ab}=$diag$(-1,+1)$
(this gauge was used in \ci{kru}).
In conformal gauge  $t$  satisfies the free equation of motion,
and we can fix the residual conformal diffeomorphism freedom
by the usual  condition
$$ t= \k \tau \ , $$
 relating the  world-sheet and the target space energies.
Then \rf{cha} implies
\be\la{chaii}
\E=  \k \ , \ \ \ \ \ \ \ \ \ \
 \J =  \inti  D_0 \a  \ . \ee
The  problem with the conformal gauge choice turns out to be related to the
fact
 that here  the energy is homogeneously distributed
along $\s$ while the  angular momentum $J$ is not, while the situation  on the spin
chain side is just the opposite.
 As a result, the comparison of the
expressions for the actions and the energies becomes
 complicated; in
particular,   the expression for $E$ in terms of $J$
contains ``non-local'' contributions
(given by  multiple integrals over  $\s$).

The equations of motion for $\a$ and $n_i$   are  found to be
\be \la{rra}  \del^p D_p\a =0  \ , \ \ \ \ \ \ \  \ \ \ \
\del^p q_p^i=0 \ , \ee
where the  latter  may be explicitly written as
\be \la {nnw}
D^p \a \ \ep_{ijk}  n_j \del_p n_k - \ha (\del^p \del_p n_i)_\perp =0
 \ ,
 \ee
 \be \la{pee}  (m_i)_\perp \equiv m_i -  (n_k m_k)n_i \ . \ee
 Equivalently,  \rf {nnw} may be written as
 \be \la{koy}
D^p \a \  \del_p n_i =- \ha  \ep_{ijk} n_j \del^p \del_p n_k  \ . \ee
 The  conformal gauge constraints
\be \la{ponn}
(\del_0 t)^2 + (\del_1 t)^2 = (q^i_0)^2+ (q^i_1)^2\ , \ \ \ \
\del_0 t \del_1 t  =  q^i_0 q^i_1 \  \ee
are easily solved  expressing  $D_p \a$ in terms of $n_i$ \foot{
 We choose  particular signs in  the solution to ensure
  regularity of large $\k$ or large $\J$ expansion.}
$
D_\pm \a = \pm \sqrt { \k^2 - \fo (\del_\pm n_i)^2}, $ where
$\del_\pm = \del_1 \pm \del_0 $,
i.e.
\be \la {ytq}
 D_0 \a =\ha \bigg[ \sqrt { \k^2 - \fo (\del_+ n_i)^2}
+ \sqrt { \k^2 - \fo (\del_+ n_i)^2}  \bigg] =
\k -  { 1 \ov 8 \k} (\del_1 n_i)^2 + \ldots\ , \ee
\be \la {ytqa}
 D_1 \a =\ha \bigg[ \sqrt { \k^2 - \fo (\del_+ n_i)^2}
- \sqrt { \k^2 - \fo (\del_+ n_i)^2}   \bigg] =
 -  { 1 \ov 4 \k} \del_1 n_i \del_0 n_i  + \ldots\ .  \ee
 We expanded in  large $\k$ which is related  to expansion
 in small $\bl = {\l \ov J^2}$. Indeed,
 \be \la{hop}
  \J  = \inti  D_0 \a
 = \k - { 1 \ov 8 \k} \inti (\del_1 n_i)^2 +  O({ 1 \ov  \k^2})
 \ ,  \ee
 i.e. (see \rf{lama}) 
 \be
 \bl = { 1 \ov \k^2}  + { 1 \ov 4\k^4 } \inti (\del_1 n_i)^2 +
  O({ 1 \ov  \k^6})
 \ . \ee
 Thus in the conformal gauge  the   natural  expansion is
  in powers of  $1\ov \k$, while on the spin chain
 side it is  the expansion in  powers of $\bl$. The two coincide at the
 leading order, but
  it is clear that beyond
 the leading order (and keeping $n_i$ general)
  the two expansions  are related in an indirect
 way  which is effectively non-local in $\s$.

 Eliminating $D_p \a$ from \rf{koy} using \rf{ytq},\rf{ytqa}
 we get an  equation for $n_i$.\foot{One can check that the
 equation for $\a$  in \rf{rra}
  is then  identically satisfied as a consequence of the
 equation for $n_i$.}
 Expanding in large $\k$ we get
 \be \la{oyy}
\k \del_0 n_i =\ha  \ep_{ijk} n_j ( \del^2_1 - \del_0^2)  n_k  +
O({1 \ov \k})  \ .
\ee
If we first assume that all derivatives stay finite  at large $\k$
then
the leading-order equation  is simply
$\del_0 n_k=0$.  Then also  $ (\d^2_1 n_i)_\perp=0$, or
$\d^2_1 n_i = - (\d_1 n)^2 n_i$.
 Multiplying this by $\d_1 n_i$ we get  also
  $(\d_1 n)^2= 4 m^2=\const$, and thus $n_i$
  has $\cos 2m\s$ and $\sin 2m\s$ as  its  two
  non-zero components. Then $ \E= \sqrt{ \J^2 + m^2}$ and
  $\Q_i =  \k \inti n_i =0$.
   This corresponds to the simplest
   circular string solution with
    two equal angular
   momenta $J_1=J_2$ (this is,  in fact,
     an exact solution of the full
  string  equations \ci{ft2}, see also below).
   Thus expanding in $\k$ in this way  corresponds to expanding near
  this special  circular  solution.

Instead, it is more  natural to follow \ci{kru} and assume that
$\k \del_0 n_i$ and $\del_1 n_i$
stay  finite in the limit, i.e. to rescale the time coordinate
(cf. \rf{resca}) 
 \be \la{joi}
\tau \to  \k  \tau \ , \ \ \ \ \ \ \ \ \
\k \del_0 \to  \del_{0 } \ , \ \ \  {\rm i.e.} \ \ \ \
t \to  \k^2 t \approx   \bl^{-1} t \ .  \ee
Here we used that $t=\k\tau$; note  that this is the same
rescaling of time $t$ that was needed in \rf{ll} or \rf{gh}
to eliminate the $\bl$ dependence.
 Then terms with more time derivatives will be suppressed
 by higher powers of $1\ov \k$.
Observing that $(\del_\pm  n_i)^2 $ in \rf{ytq},\rf{ytqa}  is
now equal to
$(\del_1  n_i \pm { 1\ov \k} \d_0 n_i)^2 $  we find that
\rf{oyy} becomes
\be \la{oy}
   \del_0 n_i =\ha  \ep_{ijk} n_j  \del^2_1  n_k  + {1\ov \k^2}
   G^{(1)}_i
 + O( {1\ov \k^4})   \ ,
    \ee
\be\la{uuo}
  G^{(1)}_i  =
- { 1 \ov 2 } \ep_{ijk}  n_j \del^2_0 n_k  -
{ 1 \ov 4 }  (\del_0 n \del_1 n) \del_1 n_i
 + \ { 1 \ov 8 } (\del_1 n \del_1 n) \del_0 n_i
 \ .   \ee
 The leading term thus
    takes the form of the LL  equation
    linear in time derivative  which is  equivalent to  \rf{ll}.
    Solutions of this leading-order equation include \ci{kru} non-trivial folded
    and circular string configurations   (which are large $\k$
    limits of the corresponding exact solutions in \ci{ft2,ft4,afrt}),
     i.e.  this is a natural starting point of an expansion
    that should describe  generic  string states with large $J$.

 Given  the remarkable fact
that this leading-order equation is {\it linear} in the time
derivative, solving  \rf{oy} perturbatively in
$1\ov \k$ one is able to eliminate
all time derivatives from the
correction terms in favor of spatial derivatives and thus to convert
\rf{oy}  into a local  equation with only spatial derivatives
  appearing on
the r.h.s., i.e. into an equation of the type expected  on
the  spin chain side (following from  \rf{expe}).

It is important to stress that this ``re-expansion'' of  the original
$R \times S^3$ \sm equations of motion effectively selects a
subclass of solutions with large $\J$ or large $\k$. That eq.
\rf{oy} does not as a result describe  all possible solutions of
  $R \times S^3$ \sm is just as well since other solutions which
  do not carry large  angular momentum
  $\J$ are not dual to SYM operators
  from the $SU(2)$ sector and thus should not be related to eigenstates
  of spin chain Hamiltonian.

  Another remark is that
this large $\k \sim \J$  conversion  of the  original second-order
equations into a first-order one
 is  similar to the usual large friction case  or
non-relativistic large mass limit.
The role of large
$\k$ or $\J= (\sqrt{\bl})^{-1} $ is indeed analogous to that of
 large \lc momentum $p^+$ in other similar cases like BMN one
 where one expands near a point-like configuration. Again, the
  novelty of the
 present expansion is that it isolates
 a sector of non-trivial extended solitonic string states
 which are far from being point-like.

 The leading term in \rf{oy}  corresponds to  the following action
 for $n_i$
\be \la{kopj}
I =  J  \int d \tau \inti \ L \ , \ \ \ \ \ \ \ \ \ \
L= C_0 - { 1 \ov 8}  ( \del_1  n_i)^2  \ ,  \ee
where  we took into account the above rescaling of $\tau$
($C_0\sim \d_\tau n$ refers to the new $\tau$)
to combine the string tension factor $\sql$ with the $\k$ factor coming
from $d \tau$ into $J$ (which is the same as $\sql \k$ to leading order
in $1\ov \k$). This action is equivalent to \rf{gh}.
The explicit form of $L$ in terms of independent  angular
coordinates in \rf{nel}  is
\be \la{ops}
L =  \cos 2 \psi \   \del_0  \b  \  -  \
 \ha \big[ (\del_1\psi)^2  + \sin^2 2 \psi  \ (\del_1\b)^2\big]  \ . \ee
  To  see the emergence of the WZ term $C_0$ \foot{Note that this
  is not the
``usual''  covariant  \sm WZ term which contains both
$\tau$ and $\s$ derivatives:
$C_0$  term is local in $\s$ (we did  not have a
$B_{mn}$-term in the original \sm action).}
 directly at the level of
  the action
we may  ``boost'' $\a$, i.e.
 introduce  instead   a new ``light-cone'' coordinate $u$
\be \la{newc}
u\equiv \a - t  \ . \ee
 Then the original \sm Lagrangian
   may be written in conformal gauge as
  \be \la{uoli}
L =  -  \del^p t D_p u  - \ha  (D_p  u )^2  -
{ 1 \ov 8}  (\del_p n_i)^2  \ . \ee
Using that $t=\k \tau$ and dropping a total derivative term we get
\be\la{pop}
 L=  \k  C_0  - \ha   (D_p u)^2  - { 1 \ov 8} (\del_p n_i)^2
  \ , \ee
 or,  after the  rescaling \rf{joi} of the time coordinate
 \be \la{koli}
 L =   C_0   - { 1 \ov 8}  (\del_1 n_i)^2  -\ha  (D_1 u )^2
  +  { 1 \ov 2 \k^2}  [(D_0 u )^2  + \fo (\del_0 n_i)^2] \ . \ee
Observing that according to \rf{ytq} (after the rescaling \rf{joi})
\be \la{kou}
D_0 u = D_0 \a -1 =
- { 1 \ov 8 }(\del_1 n_i)^2
 - {1 \ov 2\k^2} \big[{ 1 \ov 64}  (\del_1 n_i)^4  +
   \fo (\del_0 n_i)^2 \big]  + O({ 1 \ov  \k^4}) \ , \ee
  \be \la{kouu}
 D_1  u  =D_1  \a =    -  { 1 \ov 4 \k^2}{  \del_0 n_i \del_1 n_i }
+ O({ 1 \ov  \k^4})  \ , \ee
and assuming that we can use the constraints to eliminate
$u$ from the action (which, in general, requires   a  justification
but here does lead to the correct result  for the first subleading term)
we get
\be \la{poli}
 L =   C_0   - { 1 \ov 8}  (\del_1 n_i)^2
  +  { 1 \ov 8 \k^2}  [(\del_0 n_i)^2  + { 1 \ov 16}  (\del_1 n_i)^4
    ]
  + O({ 1 \ov  \k^4}) \ . \ee
The final step is to make a field redefinition to eliminate the
time derivative term from the correction. This amounts to
the use of the leading-order equation in \rf{oy}
implying
\be \la{hut} (\del_0 n_i)^2 = \fo [ (\del^2_1  n_i)^2  -
 (\del_1 n_i)^4] + O({ 1 \ov  \k^2}) \ . \ee
 As a result, we get (in terms of redefined field $n_i$)
 \be \la{loli}
 L =   C_0   - { 1 \ov 8}  (\del_1 n_i)^2
  +  { 1 \ov 32 \k^2}  [ (\del^2_1  n_i)^2  -   { 3 \ov 4} (\del_1 n_i)^4]
  + O({ 1 \ov  \k^4}) \ . \ee
  We shall obtain the same action \rf{loli}  from a more systematic
 approach of the next subsection, and later in Sections \ref{sec:expe} and \ref{sec:qc}
we  will check that the
 coefficients of the 4-derivative terms in \rf{loli} match
 the ones appearing
 in the effective action \rf{expe} of the spin chain.
Similar procedure can be applied at higher orders,
  converting the original \sm action into the first-time-derivative
  action of the type \rf{expe}.

  As already mentioned,
  the  problem of the conformal gauge we used above is that
  here the expansion goes in powers  of ${ 1 \ov  \k^2} $
  instead of powers of $ { 1 \ov  \J^2} = \bl$ and thus
  the comparison with the spin chain side is indirect
  beyond the leading
  order:
  converting the ${ 1 \ov  \k^2} $ expansion into
   $ { 1 \ov  \J^2}$ one, using the expression for $J$ \rf{hop},
    brings in
   non-local terms  given by powers of integrals over $\s$.
   For that reason we shall
   now describe an alternative gauge fixing procedure
   which is better suited  for establishing
   the correspondence between the
   spin chain and the string \sm  effective actions.

\setcounter{footnote}{0}

\subsection{Non-diagonal ``uniform'' gauge} \label{sec:nondiaggauge}

Let us start  with rewriting  generic  string \sm action in
first-order form  and then discuss  gauge fixing.
Given
\be \la{yi}
L = - \ha \sqrt{-g } g^{pq } G_{\m\n}(x) \d_p x^\m \d_q x^\n
\  \ee
and introducing the momenta
\be \la{yy}
p_\m
= -  G_{\m\n}( A \d_0 x^\m  + B \d_1 x^\m)  \ , \ \ \ \ \ \ \ \
A\equiv \sqrt{-g } g^{00 }\ , \ \ \ \
B\equiv \sqrt{-g } g^{01 }\ ,
\ee
we can rewrite $L$ in the first-order form
with respect to the time derivatives
\be \la{yyye}
L =  p_\m \del_0 x^\m  +
{ 1 \ov 2} A\inv  \big[ G^{\m\n}  p_\m  p_\n  +
 G_{\m\n}(x) \d_1 x^\m \d_1 x^\n\big]  +
{ B   A\inv } \    p_\m \d_1 x^\m
\ .  \ee
Here
 $A^{-1}$ and $BA^{-1}$ play the role of  Lagrange
multipliers for the constraints.

The action for \rf{yyye}
is reparametrization
invariant, and then as in \ci{ggrt} (in flat space)
and in \ci{mtt} (in $AdS$  space) one may fix  a gauge,
e.g., on a combination of coordinates $x^\m$ and momenta
$p_\m$.
The result is of course equivalent to the corresponding gauge
fixing directly in the Polyakov's action (cf. \ci{mtt,met})
but starting with \rf{yyye} may have some conceptual advantages
since, in the presence of  isometries, some  momenta
 are related to  densities of  conserved currents.

In the case of our interest $L$ in \rf{yi} is given by \rf{st}
and so its first-order version \rf{yyye}
has  the following explicit form
$$
L =  p_t \del_0 t   +  p_\a D_0 \a  +
 p_i \del_0 n_i  $$
\be \la{yyyi} +\   {1 \ov 2} A\inv   \big[ - p^2_t  + p_\a^2
+ 4 p^2_i
  + (\d_1 t)^2 +  (D_1 \a)^2    + \fo  (\d_1 n)^2   \big]
  \ee  $$
  + \
{ B A\inv }  (   p_t \d_1 t +   p_\a D_1 \a + p_i \d_1 n_i )
\ .  $$
Here  we have chosen to couple $p_\a$ to $D_0 \a$ and not
to $\d_0 \a$.
The equations of motion for $p_t,p_\a$ are equivalent to
the definitions of the momenta in terms of $\del_q x^\m$ and $A,B$ in
\rf{yy},
\be \la{pep}
p_t =  A \d_0 t + B \d_1 t \ ,  \ \ \ \ \ \
p_\a = - A D_0 \a - B D_1 \a  \ . \ee
 Note also that we have
  defined   $p_i$ so that
 they are different from the  usual canonical momenta for $n_i$:
  $p_i$ does not include the contribution of
   the   $\del_0 n_i$-dependence  of $C_0$ in
$D_q\a$, i.e.
\be \la{pip}
p_i = - \fo ( A \d_0 n_i + B \d_1 n_i ) \ .  \ee
The equations  for the metric components
 $A$ and $B$ give the  constraints
\be\la{ions}
- p^2_t  + p_\a^2  + 4 p^2_i
  + (\d_1 t)^2 +  (D_1 \a)^2    + \fo  (\d_1 n)^2 =0 \ , \ee
 \be\la{jons}
  p_t \d_1 t +   p_\a D_1 \a + p_i \d_1 n_i =0 \ . \ee
The equations  for $t, \ \a$ and $n_i$ are found to be
\be \la{juo}
\d_0 p_t  + \d_1 [A^{-1} ( \d_1 t  + B  p_t) ]    =0 \ , \ee
\be \la{kuo}
\d_0 p_\a  + \d_1[ A^{-1} (D_1 \a   + B  p_\a) ]    =0 \ , \ee
\be \la{puo}
- \ha \ep_{ijk} n_j \big[
p_\a  \del_0 n_k + A^{-1}  (D_1 \a   + B p_\a )  \del_1 n_k
\big]
+  \ \bigg( \d_0 p_i   +   \d_1 [A^{-1} ( \fo  \d_1 n_i   + B  p_i )]
\bigg)_\perp
 =0 \ .  \    \ee
 $n^2_i=1$ implies that $n_i \del_q n_i=0,\
p_i n_i=0$ and that the  variation over $n_i$
should be orthogonal to $n_i$.
The first
$\ep_{ijk}$-term  in \rf{puo} comes from the variation
of $C_q$ in $D_q \a$
(an additional gradient variation term is proportional
to the equation \rf{kuo} for $\a$ and thus is ignored).

The conserved charges  corresponding to  the
invariances under translation in time,
translation in $\a$ and  $O(3)$ rotations
($\delta n_i = \ep_{ijk} \b_j n_k$) of $n_i$  are
(see  \rf{cha},\rf{kop},\rf{okk})
\be \la{chai}
\E \equiv P^t= \inti\ \cH \ , \ \ \ \ \
\J = \inti\ p_\a  \ , \ \ \ \ \ \ \
\Q_i = \inti\ q_i \ , \ \ee
\be\la{quf}  \cH \equiv -  p_t \ , \ \ \ \ \ \ \ \ \ \ \ \ \
q_i = p_\a  n_i  + 2 \ep_{ijk} n_j p_k  \ . \ee
Let us now choose a gauge.
The  gauge used in the previous section
was   the orthogonal one
 $A=1, B=0$ and  $t=\k\tau$.
Having in mind comparison
with the spin chain  it is natural to request that
translations in time in the target space and on the
world sheet should indeed
be related. Also, we should ensure that  the angular momentum $\J$
 is
homogeneously distributed along the string,
i.e. let us require\foot{This is another
example  illustrating that  in curved space it is often natural to use
a non-conformal gauge. In fact, use of such a more general  gauge
(or a  special choice of $\tau$ and $\s$ variables)
may be ``required'' by duality to gauge theory.
Again, it  is true of course
that all final physical observables (like values of energies  on particular
solutions) should be gauge-independent. But an ``off-shell' comparison
of string and gauge theory may be greatly facilitated by the
right choice of world-sheet coordinates.}
\be \la{gaug}
(i) \ \ t= \tau \ , \ \ \ \ \ \ \ \
(ii)\ \  p_\a = \J =\const \ . \ee
Here we have chosen to set a proportionality coefficient between $t$ and
$\tau$  to 1 but later we will
 need to rescale $\tau$ as in the above conformal gauge
discussion.
In this case \rf{pep} and \rf{pip}  imply
\be \la{imm}
A=- \cH \ , \ \ \ \ \  \J= \cH D_0 \a - B D_1 \a \ , \ \ \ \ \
p_i= \fo ( \cH \d_0 n_i - B \d_1 n_i) \ ,  \ee
and the constraints \rf{jons},\rf{ions} give
\be\la{fi}
D_1 \a = - { 1 \ov \J} p_i \d_1 n_i\ , \ee
\be\la{dons}
\cH\equiv \cH(n,p)=  \sqrt{ \J^2   + 4 p^2_i
  +  { 1 \ov \J^2 } (p_i \d_1 n_i)^2
     + \fo  (\d_1 n)^2 }  \ .  \ee
Using the constraints, the resulting ``reduced''
Lagrangian for the
independent  variables $n_i$ and $p_i$ is  found
from \rf{yyyi} to be
$
L= \J D_0 \a + p_i \del_0 n_i - \cH . $
Omitting the total derivative $\J \d_0 \a$ term, we get
\be\la{kiop}
L(n, p)  = \J C_0 + p_i \del_0 n_i -  \cH(n,p)  \ ,  \ee
where $C_0$ is again the same WZ term as in \rf{pop} or \rf{cce}.
Thus both the   first and the second
terms here  depend on $\d_0 n_i$,
 reflecting the fact that $p_i$ was defined not to include a
 contribution from $C_0$ in   $D_0 \a$.

 Next, we  should  check that this Lagrangian  \rf{kiop} does indeed
 lead to the correct equations for $n_i,p_i$, i.e.
 that here it is legitimate to use the gauge conditions and
 the constraints directly in the action.
With the above gauge choice the general equations
\rf{juo}--\rf{puo} become
\be \la{ku}\del_0 \cH  -   \d_1  B =0 \ , \ \ \ \ \ \
 \d_1[ \cH\inv  ( B   - { 1 \ov \J^2} p_i \d_1 n_i) ]    =0 \ , \ee
\be \la{wpuo}
- \ha  \J \ep_{ijk} n_j \bigg[
  \del_0 n_k -\cH^{-1}
 (B  - { 1 \ov \J^2} p_i \d_1 n_i)  \del_1 n_k\bigg]
+  \ \bigg( \d_0 p_i   -   \d_1 [\cH^{-1} ( \fo  \d_1 n_i
 + B  p_i )]
\bigg)_\perp
 =0 \ ,   \    \ee
 where according to \rf{imm}
 $p_i= \fo ( \cH \d_0 n_i - B \d_1 n_i)$.
 We are to compare \rf{puo},\rf{pip}  with the equations  for $n_i$ and $p_i$
that follow directly from \rf{kiop}
\be \la{juy}
\del_0 n_i -\cH \inv  [ 4 p_i + {1\ov \J^2} (p\del_1 n) \d_1 n_i ]
=0 \ , \ee
\be \la{kpd}
- \ha \J \ep_{ijk} n_j \del_0 n_k  + \bigg( \d_0 p_i
-  \del_1 \big[\cH\inv  ( \fo \del_1 n_i +
 {1\ov \J^2} (p\del_1 n) p_i\big]\bigg)_\perp  =0 \ . \ee
We conclude that they agree  provided
\be \la{ghi}
B= {1\ov \J^2} (p\del_1 n)\ , \ \ \ \ \  {\rm i.e. }
\ \ \ \  \  D_1 \a = - \J B \ . \ee
We are still to check that this  does not contradict
 the first equation in
\rf{ku}, i.e.
\be \la{enq}
\del_0 \cH  =  {1\ov \J^2} \del_1 (p\del_1 n) \ . \ee
 This, indeed,  follows from the
 equations of motion \rf{juy},\rf{kpd}
 for $p_i$ and $n_i$.\foot{
 In general, given the Lagrangian which is first-order in
 $\del_0 n_i$ one finds that (omitting $\perp$ sign)
 $\d_0 p_i - \ha \J\ep_{ijk} n_j \d_0 n_k =
 \d_1 {\del \cH \ov \del \del_1 n_i}  ,$ \
 $\d_0 n_i =  {\del \cH \ov \del p_i}$,
 from where it follows that (WZ term does not contribute)
 $$
 \d_0 \cH
 = {\del \cH \ov \del \del_1 n} \d_0 \del_1  n +
 {\del \cH \ov \del p} \d_0 p
 = \big(  - \d_1 {\del \cH \ov \del \del_1 n} \d_0   n  +
 {\del \cH \ov \del p} \d_0 p \big) +
   \d_1 \big( {\del \cH \ov \del \del_1 n} \d_0   n ) \ . \ \ \  $$
   The first bracket vanishes on the equations of motion,
   so $\cH$ is  conserved.
   Computing $ {\del \cH \ov \del \del_1 n} \d_0   n $
    using \rf{juy} one finds indeed that it is equal to $B$, i.e. to
   $ {1\ov \J^2} (p\del_1 n).$}
 Thus  our procedure is consistent: the extra
 condition \rf{ghi}
 is  a  solution of our equations.
 We could, in fact,   impose \rf{ghi} and $p_\a =\J$ in \rf{gaug}
 as   two reparametrization gauge conditions; then an extra
  choice of
 $t=\tau$ in \rf{gaug}   would   correspond to
  fixing a residual reparametrization freedom, as in the
 conformal gauge.

 Since $p_i$ enters \rf{kiop} only algebraically, we can
 now solve for it  and get a (non-polynomial)
  action that depends only on $n_i$ and
 its first derivatives $\d_0 n_i$ and $\d_1 n_i$.  We find from \rf{juy}
\be\la{po}
p_i = \fo \cH [ \d_0 n_i  - { (\d_0 n \d_1 n) \ov 4 \J^2 + (\d_1 n )^2 }
\del_1 n_i ] \ , \ \ \ \ \ \ \    p_i n_i =0 \ , \ee
\be \la{trq}
p_i \del_1 n_i =\fo \cH { (\d_0 n \d_1 n) \ov 1 + { 1 \ov 4 \J^2} (\d_1 n )^2
}=  \J^2 B \ .
\ee
Substituting  $p_i$  into $\cH$ gives
\be\la{enn}
\cH = \J  { 1+ { 1 \ov 4 \J^2 } (\d_1 n )^2 \ov
\sqrt{ [1+ { 1 \ov 4 \J^2 } (\d_1 n )^2][ 1 - { 1 \ov 4  } (\d_0 n )^2]
+ { 1 \ov 16  \J^2 } (\d_0 n \d_1 n )^2}}
\ ,  \ee
and  eliminating  $p_i$ from the Lagrangian \rf{kiop},
we get\foot{We also find that
$D_0 \a=  { 1 - { 1 \ov 4  } (\d_0 n )^2
\ov  \sqrt{ 1+ { 1 \ov 4 \J^2 } (\d_1 n )^2  - { 1 \ov 4  } (\d_0 n )^2
+ { 1 \ov 16  \J^2 }[  (\d_0 n \d_1 n )^2   - (\d_1 n )^2
(\d_0 n )^2 ]  }} $.}
\be \la{final}
L(n ) = \J C_0  - \cH_n \ , \ \ \ \ \ \ \ \ \
 \cH_n \equiv  \cH - p_i \d_0 n_i \ ,
\ee
\be \la{konn}
\cH_n
= \J \sqrt{ [1+ { 1 \ov 4 \J^2 } (\d_1 n )^2][ 1 - { 1 \ov 4  } (\d_0 n )^2]
+ { 1 \ov 16  \J^2 } (\d_0 n \d_1 n )^2}
\ ,  \ee
Equivalently,
\be \la{klpo}
\cH_n =  \sqrt{ - \det \g_{pq} } \ , \ \ \ \
\g_{pq} = \td \eta_{pq} + \fo \del_p n_i \del_q n_i \ , \ \ \ \
\td \eta_{ab} \equiv  {\rm diag} ( -1, \J^2) \ .  \ee
Remarkably, if not for the WZ term $C_0$, the expression for
\rf{final}    is reminiscent of the
Nambu Lagrangian in a static gauge (suggesting that there may be
 a more direct way of deriving this action).

We have thus managed to recast the original \sm  action in
the form  resembling \rf{expe}. The remaining steps are:

 (i)
to define  a consistent $1\ov \J$ expansion,  and then

(ii) to
eliminate time derivatives in $\cH_n$ by field redefinitions
 order by order in $1\ov \J$.

\noindent
To define  the large $\J$ expansion near the same ``classical ferromagnet''
limit as in the conformal gauge (cf. \rf{loli}) we need
to rescale (cf. \rf{resca},\rf{joi})
\be\la{resc}
\tau \to \J^2 \tau = { \bl^{-1}} \tau\ , \ \ \ \ {\rm i.e.}\ \ \
\ t \to   \bl^{-1} t \ , \ \  \ \ \ \ \
\del_0 \to { 1 \ov \J^2} \del_0 \ , \ee
thus getting the string action in the form
\be \la{bbb}
I = J \int d\tau \inti\  L \  , \ \ \ \ \ \ \
L=  C_0 -  \bar  \cH \ , \ \ \ \ \ \ \bar  \cH \equiv \J \cH_n  \ee
\be \la{lonn}
 \bar \cH
=  \J^2 \sqrt{ [1+ { 1 \ov 4 \J^2  } (\d_1 n )^2][ 1 - { 1 \ov 4 \J^4 } (\d_0
n )^2]
+ { 1 \ov 16  \J^6 } (\d_0 n \d_1 n )^2}
\ ,  \ee
i.e. (omitting constant term  in $\bar \cH$)
$$L= C_0 - { 1 \ov 8  } (\d_1 n_i)^2  +  {1 \ov 8 \J^2} \big[(\del_0
n_i)^2   +  { 1 \ov 16}
(\del_1 n_i)^4  \big] $$
\be \la{llii}- \  {1 \ov   32 \J^4} \big[
  (\del_0 n_i \del_1 n_i)^2
-{1 \ov 2}   (\del_0 n_i)^2  (\del_1 n_k)^2
+ { 1 \ov 32 } (\del_1 n_i \del_1 n_i )^3  \big]
+ O({1 \ov \J^6}) \ .  \ee
The second and fourth derivative terms here are the same as in
 the conformal gauge expression  \rf{poli}, and thus the equations
 of motion  to this order
 also have the same form \rf{oy},\rf{uuo}.

Note that it is the energy density \rf{enn}, or, in terms of redefined
time derivative
\be\la{kenn}
\cH = \J  { 1+ { 1 \ov 4 \J^2 } (\d_1 n )^2 \ov
\sqrt{ [1+ { 1 \ov 4 \J^2 } (\d_1 n )^2][ 1 - { 1 \ov 4 \J^4 } (\d_0 n )^2]
+ { 1 \ov 16  \J^6 } (\d_0 n \d_1 n )^2}}
\   \ee
\be\la{eedd}
  =\J + \J^{-1} \bigg[
     { 1 \ov 8  } (\d_1 n )^2  +  {1 \ov 8 \J^2}[(\del_0 n_i)^2
  -   { 1 \ov 16  }  (\del_1 n_i)^4]
 + O({1 \ov \J^4})\bigg]  \   \ee
that is conserved on the equations of motion.
It is only after a  field redefinition  $n_i\to \td n_i$
 that eliminates
all time derivatives in the $\bar \cH$ part of $L$  in \rf{llii},
i.e. that puts $L$ into the form
\be \la{exep}
L= C_0(\td n) - H(\d_1 \td  n, \d_1^2 \td n, \d^3_1  \td  n,\ldots) \ ,
\ee
that  $H$ should be conserved.\foot{
Indeed,
if $H$ depends only on spatial derivatives
  the equations of motion following from  take  the form
 $
 \ha  \ep_{ijk} n_j \d_0 n_k =  (\d_1[  { \del H \ov \del \d_1 n_i}
 + \d_1 { \del H \ov \del \d^2_1 n_i } + \ldots])_\perp \equiv K_i,
 $
and  then $ K_i \d_0 n_i  =0$ so that
 $$
\d_0 H = \d_0 \d_1 n_i  { \del H \ov \del \d_1 n_i }
 +  \d_0 \d_1^2 n_i  { \del H \ov \del \d_1^2 n_i} + \ldots =
 \d_1 \big( \d_0 n { \del H \ov \del \d_1 n_i}
 +  \d_0 \d_1 n_i { \del H \ov \del \d_1^2  n_i} + \ldots\big) \ ,
$$
 and  the integral of $\d_0 H$ vanishes. }
  On general grounds, since (the integral of) $\cH$ generates
  translations  in $t$ and (the integral of) $H $ generates
  translations  in $\tau$, and $t \sim  \tau$ according to the
  above choice, the two   must be equivalent  on-shell
  (after the same  field redefinition done in $\cH$,
   modulo a total  $\d_1$ derivative), i.e.
   \be \la{[}
[\cH(\d_0 n, \d_1 n)]_{_{\rm field\  redefn,\  on-shell}}
=  \J^{-1} H( \d_1 \td n, \d^2_1 \td n, \ldots)  + \del_1 F\ . \ee
This is  what is required for  correspondence with
the spin chain results.
In contrast to the conformal gauge case, the relation between the
space-time and 2-d energies here
does   not involve non-local (multiple $\s$
integral) terms.

 The form of   field redefinition which
 is needed to put  $L$ in \rf{bbb}
   into the form \rf{exep} is suggested by the comparison of the $O(3)$
   conserved charges which should also match: for  \rf{exep}  we get instead
   of the expression in
   \rf{cha},\rf{quf}
  \be  \S_i = \J \inti\  \td n_i \ ,   \ee
  implying that the required field redefinition  should be
  (we rescale $q_i$ in \rf{quf} by factor of $\J$)\foot{Once  the action is put into the form
\rf{exep}, the density of the $O(3)$ rotational current is determined
simply
by the WZ term $C_0$, i.e. it should be equal (up to a total
derivative)
 to $\td n_i$.
}
  \be \la{hp}
  \td n_i =  q_i(n,\d_0 n,\d_1 n) + \d_1 f =
  n_i + O({1 \ov \J^2})\ , \ee
  where a  total derivative term $\d_1 f$ may be needed
  to ensure that $\td n_i^2=1$.

Let us demonstrate how this works   to the two leading orders
in large $\J$ expansion (see also Appendix A).
Starting with \rf{llii}
let us do a field redefinition that  converts the
$(\del_0 n_i)^2$ term into its ``on-shell'' value \rf{hut}, i.e.
$
[(\del_0 n_i)^2]_{\rm on-shell}= \fo ((\del^2_1 n)_\perp)^2
= \fo[ (\del^2_1 n)^2 -  (\del_1 n)^4 ] . $
Observing that the variation of the first two leading terms
 in \rf{llii} is proportional to the leading-order equations
 (omitting total spatial derivative)
 $ \delta [C_0 -  { 1 \ov 8  } (\d_1 n )^2 ]
=  \big[ \ha \ep_{ijk} n_j
\del_0 n_k  + \fo (\del^2_1 n_i)_\perp\big]\ \delta n_i  ,  $
we conclude that we need
\be \la{poll} \delta n_i = { 1 \ov 2 \J^2} [ - \ha \ep_{ijk} n_j
\del_0 n_k  + \fo (\del^2_1 n_i)_\perp] \ , \ \ \ \ \ \
[\delta n_i]_{_{\rm on-shell}}=  { 1 \ov 4 \J^2} (\del^2_1 n_i)_\perp \ .
 \ee
This redefinition effectively replaces
${1 \ov 8 \J^2}(\del_0 n_i)^2$ term
  with  ${1 \ov 32 \J^2}((\del^2_1 n_i)_\perp)^2$,
so that in terms of redefined field $\td n_i = n_i - \delta n_i$ we get
\rf{exep} with (omitting tilda on $n_i$)
\be \la{kuru}
H = H_0 + H_1 +  H_2 + O(\bl^3)
\ , \ \ \ \ \ \ \ \ \
H_0=   { 1 \ov 8  } (\d_1  n )^2\ , \ \ee
\be \la{kur}
H_1 =   -
{\bl \ov 32 } [ (\del^2_1  n_i)^2   -  {3\ov 4} (\del_1  n_i)^4 ]
   \ ,   \ee
    where we used \rf{lama} to express $\J$ in terms of $\bl$. 
This  happens to be the same result as in the conformal
 gauge \rf{loli}  with $\k \to \J$.
 Since the two actions -- \rf{llii} and \rf{exep} --
are related by a field redefinition,
they have equivalent equations of motion, i.e. \rf{exep} with $H$
given by \rf{kur}
is guaranteed to reproduce the same  string solutions.

The field redefinition we have used is indeed equivalent, on-shell,
and to the leading order in $1\ov \J$, to the one in \rf{hp}
that transforms $n_i$ into the charge density $q_i$ 
\be \la{qn}
n_i = q_i - { 1\ov 2 \J^2}  \ep_{ijk} q_j \d_0 q_k + O({1\ov \J^4}) \
\ \to \ \
q_i + { 1 \ov 4 \J^2} (\del^2_1 q_i)_\perp + O({1\ov \J^4})\ , \ \ \ \
\td n_i \equiv q_i \ . \ee
Let us now show that the  energy density \rf{kenn} becomes indeed
equal (up to an overall  factor of $\J$)
 to $H$ in \rf{kur}
after the above field redefinition and evaluation of the result
on  the equations of motion.
 It is important to stress
that while the use of  equations of motion
in the  action  is equivalent to field redefinitions
(at least, to leading order), this is not so in the energy:
field redefinition and evaluation on-shell are two different steps.
 Notice that compared to $\bar \cH$ in \rf{llii}
 the sign of the $(\del_0 n_i)^2$ term in $\cH$  is the  opposite,
 so if we would simply
 evaluate $ \cH $ on-shell we would not match $H$ in
 \rf{kur}. Instead,  we are instructed   to do the  field redefinition
 \rf{poll} first, and it
 gives $(\d_1 n )^2 \to - \del_1^2 n_i \delta n_i .$
 Evaluating the result on the leading-order equations of motion
  we end up with (up to a total derivative, subtracting
   the constant term
  $\J$,  rescaling by factor of $\J$ and omitting 
  tilda on $n_i$)
 $$
\J (\cH - \J) =  { 1 \ov 8  } (\d_1   n )^2  - { 1 \ov 8 \J^2}
\big[
 2 \del_1^2 n_i (\del_1^2 n_i)_\perp  +
 (\del_0  n_i)^2
  -   { 1 \ov 16  }  (\del_1  n_i)^4\big]
 + O({1 \ov \J^4}) $$
 \be\la{ekdd}
 = { 1 \ov 8  } (\d_1   n )^2  -
 { \bl \ov 32} \big[(\del_1^2  n_i)^2   -   { 3\ov 4 }
  (\del_1  n_i)^4 \big]
  + O({\bl^2})\ . 
 \ee
 The role of the field redefinition is thus to invert the
 sign of the $(\d_0 n)^2$ term in $\cH$; then, upon
 evaluation on-shell, $\cH$  matches   $H$ in \rf{kur}.

It is clear that the same field
redefinition procedure should apply at higher orders of
expansion in $1\ov \J$. We demonstrate this explicitly
at the next  $1\ov \J^4$ order in Appendix A.
Here we will just quote the final result for the corresponding
term in $H$ in \rf{exep},\rf{kuru} (again omitting tilda on $n_i$
and using \rf{lama})
\foot{The same expression for the 6-derivative term in $H$ 
was found by A. Dymarsky and I. Klebanov 
by reconstructing  the $C_0-H$ Lagrangian 
from the condition that it correctly reproduces the energies 
of the folded \ci{ft4} and circular \ci{art} solutions 
expanded to the given order in $\bl$.} 
\comment{NEW}
\be\la{poj}
 H_2=
{\bl^2 \ov   64 } \bigg[ (\d^3_1 n) ^2
 -  { 7 \ov 4} (\d_1 n)^2 (\d_1^2 n )^2
  - { 25\ov 2} (\d_1 n \d_1^2 n )^2  +  {13\ov 16} 
(\d_1 n)^6
\bigg]  \ . \ee
 Below we shall compare these results with the corresponding
 expressions on the spin chain side \rf{expe}.

\comment{NEW}
It would be interesting to determine 
terms in $H$, to a given 
order in an expansion in powers of $n_i$, but to all 
orders in spatial derivatives. 
This is  possible to do for all quadratic terms
by expanding near the ``vacuum'' configuration 
$n_i = (0,0,1)$ which essentially corresponds to the BMN limit
(small fluctuations above the BPS vacuum). 
Then starting with \rf{lonn} where all higher than quadratic terms 
should be omitted  
and solving for the frequency in terms 
of the spatial momentum one finds that all quadratic terms in the effective Hamiltonian 
can be written as follows 
(modulo integration by parts) 
\be \la{quad}
H = \fo \bl^{-1}   
n_i \left( \sqrt{1 - \bl 
\d_1^2 } -1 \right) n_i  + O(n^4) 
\ . \ee
The expansion of the BMN square root here 
is in agreement with the coefficients of the quadratic terms
in \rf{kuru},\rf{kur},\rf{poj}.

%
\subsection{Some special solutions} \label{sec:sol1}

Let us  mention  how some  known two-spin solutions
of the $R\times S^3$ \sm fit into the above discussion.
A class of folded and circular string solutions with
non-constant radii \ci{ft4,afrt}
 (having $\vp_k= w_k \tau, \ \psi= \psi(\s)$ in \rf{fre})
have $n_i$ \rf{ne} satisfying
\be \la{spec}
\d_0 n_i \d_1 n_i=0 \ . \ee
As follows from \rf{lonn},\rf{kenn} in this case\foot{Here
 $p_i= \fo \cH \d_0 n_i  , \ p_i \d_1 n_i =0$ and
 $B$ and $D_1 \a$  vanish so the ``uniform'' gauge is diagonal
   (see \rf{pip},\rf{ghi}).
 Also, $D_0 \a$ is conserved,
 i.e. it depends only on $\s$, so  the ``uniform''
  gauge may be reached from the
 conformal gauge by an additional redefinition of $\s$.}
\be\la{nn}
\cH = \J  \sqrt{ { 1+ { 1 \ov 4 \J^2 } (\d_1 n )^2 \ov
 1 - { 1 \ov 4 \J^4 } (\d_0 n)^2  }    } \ , \ \ \ \ \ \
\bar \cH
= \J^2 \sqrt{ [1+ { 1 \ov 4 \J^2 } (\d_1 n )^2][ 1 -
{ 1 \ov 4\J^4  } (\d_0 n)^2]}
\ .  \ee
Another class of solutions are circular solutions
with constant radii  \ci{art}
for which the angles in \rf{fre} are
$\psi=\psi_0=\const \ , \vp_k= m_k \s + w_k \tau,$
(i.e. $\b= \ha (m_1-m_2) \s + \ha (w_1 - w_2) \tau$),
 where (in conformal gauge)
 $w_k= \sqrt{ m^2_k + \nu^2} , \
\E^2 = 2 (w_1 \J_1 + w_2 \J_2) - \n^2 $ and $\nu$ is a solution of
${\J_1 \ov w_1} + {\J_2 \ov w_2}=1, \ m_1 \J_1 + m_2 \J_2=0$.
These are homogeneous solutions: all invariants built out of
 $n_i$, in particular,  $\d_0 n_i \d_1 n_i$ are constant.

  In the special case
of circular solution with equal spins \ci{ft2}
($\J_1=\J_2= \ha \J$, $m_1=-m_2=m$)
one has $\psi_0={\pi\ov 4}, \ \b= m\s$, $
n_i= (\cos 2 m\s, \sin 2 m\s, 0)$  and   $\d_0 n_i =0$.
This is, in fact, the general {\it static }
solution of the leading-order LL equations
\rf{ll} or \rf{oy}
\be  \la{pq}\d_0 n_i =0 \ , \ \ \ \ \ (\d_1^2 n_i)_\perp =0 \ , \ \ \ \
(\d_1 n )^2=\const = 4 m^2 \ , \ee
and  is also  an exact solution of the full system to all orders in $1\ov
\J$. Here we get $\Q_i=0$ and
\be \la{yq}
\cH   = \J^{-1} \bar \cH =
   \sqrt{\J^2 + {\textstyle { 1 \ov 4 }} (\d_1 n )^2 } =
  \sqrt{\J^2 + m^2 }
 \ .  \ee
\comment{NEW}
It is easy to check that eqs.\rf{kuru},\rf{kur},\rf{poj} are
indeed in agreement 
with the expansion of \rf{yq}.
We shall also  check the 
second-order expression \rf{kur} 
against the energy of the folded string solution \ci{ft4}
in section 5.


\setcounter{equation}{0}
\setcounter{footnote}{0}

\section{Expectation value
of dilatation operator in coherent state and  ``naive''
continuum limit} \label{sec:expe}

Let us now  consider the higher order (higher-loop) corrections \rf{xpe}
on  the SYM, i.e. the spin chain,  side.
In general, one is supposed to find eigenvalues of the 
SYM dilatation operator
and compare them to  the
$AdS_5$  energy of the
corresponding string states. In the large $J$ limit \rf{lim},
this problem happened to be  essentially
semiclassical at   the leading order in $\l$
 (cf. \rf{gh}).
One may expect that the same may be true also at higher loop orders.
In this case to compare to string theory
it would be sufficient to know the analogue of the action \rf{kh},\rf{gh}
in the case 
when  higher order corrections are included in the spin chain Hamiltonian
\rf{hei}.
Our first task, therefore,
 will be to compute the action \rf{kh} that appears in
the coherent state path integral of the quantum spin chain
with the Hamiltonian given by
 the 3-loop SYM dilatation operator in the
$SU(2)$ sector.
We will then address the issue of  taking continuum limit and whether
one is actually able to ignore quantum corrections beyond the leading
(one-loop) order.
We  will  describe how to  consistently include quantum corrections
in the next section.


The one \ci{mz1}, two \ci{bks} and three \ci{bks,bei,beit}
loop
dilatation operator
of the $\cal N$=4  SYM theory  in the $SU(2)$ (2 chiral scalar operator) sector
has the form
\be \la{dod}
 D=
\sum_{r=0}^\infty { \l^r \ov (4 \pi)^{2r}}   D_{2r}  \ , \ \ \ \ \
\ \ \ \ \ \ \ \ \ \
 D_{2r} =   \sum^{J}_{a=1} \D_{2r}(a) \ ,
\ee
\be \la{ii}
\D_0 = I\ , \ \ \ \ \ \ \ \ \ \ \ \ \D_2= 2(I - P_{a,a+1}) \ , \ \ \ \ \ \ee \be
\D_4= -8 I  + 12 P_{a,a+1} - 2 ( P_{a,a+1}P_{a+1,a+2} +
P_{a+1,a+2}P_{a,a+1})  \ , \ee
$$
 \D_6 = 60 I -104  P_{a,a+1}  + 24 ( P_{a,a+1}P_{a+1,a+2} +
 P_{a+1,a+2}P_{a,a+1}) $$ \be \la{klp}
+\  4 P_{a,a+1}P_{a+2,a+3}   - 4 (  P_{a,a+1}  P_{a,a+1}P_{a+2,a+3}
+   P_{a+2,a+3}  P_{a,a+1}P_{a,a+1})                \ . \ee
Here the projection operator is 
\be \la{iiy}
P_{a,b} = \ha ( I + 4 S_a \cdot S_{b}) \ , \ \ \ \ \ \ \
S^i = \ha \s^i  \ ,  \ee
and  $\s^i$ are Pauli matrices.
Since the 3-loop term $D_6$  was not yet explicitly  derived
from SYM theory but was fixed using indirect (based, in particular, on 
integrability \ci{bks}
or superconformal symmetry \ci{beit}) considerations, let us mention that
$\D_6$ in \rf{klp}
 is  the $\a_1=0$ member of a one-parameter class of operators
in \ci{bei} (we set $\a_2$ in \ci{bei} to be zero) 
that all have similar properties like a 
consistent  BMN limit: \foot{We correct 
a misprint  in eq.(8) of \ci{bei}: the  sign of the 
$14 \alpha_1$ term there should be minus (we thank N. Beisert for
pointing this out to us).
Strangely, the  $\a_1=2$ operator  (apparently 
not consistent with the integrability) 
 appears to be 
special in what concerns comparison to string theory, see discussion in \ci{bei}.}
$$
 \D_6(\a_1)  = (60 + 6 \a_1)  I -(104 + 14 \a_1 )   P_{a,a+1}
 $$ $$ + \
 (24 + 2 \a_1) ( P_{a,a+1}P_{a+1,a+2}  +
 P_{a+1,a+2}P_{a,a+1})
+  (4 + 6 \a_1)  P_{a,a+1}P_{a+2,a+3}  $$
$$
 -\  4 (  P_{a,a+1}  P_{a,a+1}P_{a+2,a+3}
+   P_{a+2,a+3}  P_{a,a+1}P_{a,a+1}) $$
\be \la{alta} - \ \a_1   (  P_{a,a+1}  P_{a+2,a+3} P_{a+1,a+2}
+   P_{a+1,a+2}  P_{a,a+1}P_{a+2,a+3})
     \ . \ee
Using that for each  site index $a$ one has 
$
 \s^i_a \s^j_a =  \delta^{ij} + i \ep^{ijk} \s^k_a  $
it is straightforward to show that $\D_2,\D_4,\D_6$ can be written
 in the following equivalent forms\foot{Here we  use that 
 for the case of a periodic spin chain 
  which we  are  interested in
  one can shift  the summation index and thus
combine terms which are equivalent under the summation.}
\be \la{iy}
 \D_2= 2 Q_{a,a+1} \ , \ \ \ \
\D_4=  - 2( 4 Q_{a,a+1}  -Q_{a,a+2})   \ , \ee
\be\la{ity}
 \D_6 =4( 15Q_{a,a+1}    -6  Q_{a,a+2}  +  Q_{a,a+3 })
+  4 (Q_{a,a+2} Q_{a+1,a+3}   -  Q_{a,a+3} Q_{a+1,a+2})     \ , \ee
where
\be  Q_{a,b} \equiv  I - P_{a,b} = \ha ( 1 - 4  S_a \cdot S_{b})
= \ha ( 1 -   \s_a \cdot \s_{b})  \ .  \label{Qdef}\ee
In general, for \rf{alta} we get instead of \rf{ity}
$$
 \D_6(\a_1)  =4(15Q_{a,a+1}    -6  Q_{a,a+2}  +  Q_{a,a+3 })
 $$
\be\la{yity}+ \ ( 4 -  \a_1) ( Q_{a,a+2} Q_{a+1,a+3}   -
 Q_{a,a+3} Q_{a+1,a+2}) +  {5} \a_1
 Q_{a,a+1} Q_{a+2,a+3}       \ . \ee
Note that $D$ \rf{dod} interpreted as a
 generalized  spin  chain  Hamiltonian has a ferromagnetic nature:
the coefficient multiplying each linear $Q_{a,a+c}=
 \fo ( \s_a -\s_{a+c})^2$ term is positive
  (assuming that perturbation theory in $\l < 1$ applies).

Consider now a coherent state $\ket{n_a}$ defined as
\beq
\ket{\vn_a} = R(\vn_a) \ket{\uparrow} ,
\label{chs}
\eeq
where $\ket{\uparrow}$ is a spin up state and $R(n_a)$ is the rotation 
\beq
R(n)  = e^{-i\phi S_3 } e^{-i\theta S_2 }\ , \ \ \ \ \ \ \ \ \ 
n_i = (\sin \theta \ \cos \phi;\   \sin\theta \ \sin \phi;\
\cos \theta) \ ,   
\label{rot}
\eeq
where we use $\theta$ and $\phi$ (instead of $2\psi$ and $2\beta$ 
in \rf{nel})  as polar coordinates for the unit vector 
$n$. From such coherent states at each site we can built 
 a ``coherent'' state  of the whole chain as
\foot{Ironically, instead of ``coherent'', a 
better name for such state would be  ``decoherent'' 
 as it is simply a direct
 product of independent states at each site.}
$\ket{n} \equiv \prod_{a=1}^J \ket{n_a}=  
 \ket{n_1,\ldots,n_a,\ldots}$  in which case 
\be \la{ety}
\bra{n}S^i_a \ket{n} = \ha  n^i_a   \ , \ \ \  \ \ \ \ \ \ \  
n_a \cdot n_a =1 \ , \ee
\be  \la{rty}
\bra{n}Q_{a,b} \ket{n} = \ha N_{a,b} \ , \ \ \ \ \ \ \ \ \ \ 
N_{a,b}\equiv  1 - n_a \cdot n_b
  = \ha  ( n_a - n_b)^2 \ .
 \ee
We then  end up  with
\be \la{wty}
\bra{n}\D_2\ket{n} = N_{a,a+1} 
\ , \ \ \ \ \ \ \ \ \ \ \ \ 
\bra{n}\D_4\ket{n} = -4 N_{a,a+1}  + N_{a,a+2} 
 \ , \ee
\be  \la{woty}
\bra{n}\D_6\ket{n} =
30  N_{a,a+1}     - 12  N_{a,a+2}   +  2 N_{a,a+3} 
+\ N_{a,a+2}N_{a+1,a+3} - N_{a,a+3}N_{a+1,a+2}  \ . \ee
Now,  $ \bra{n}\rH\ket{n} $ in \rf{kh} should be replaced by
\be
\bra{n}\rH\ket{n} = \sum_{a=1}^J \bigg[ { \l \ov (4 \pi)^2}  \bra{n} \D_{2}(a)\ket{n}
+ { \l^2 \ov (4 \pi)^4} \bra{n}  \D_{4}(a)\ket{n}
+ { \l^3 \ov (4 \pi)^6}  \bra{n} \D_{6}(a)\ket{n}  + \ldots \bigg]\ . \ee
Next, let us take a  continuum limit as in \rf{gh} 
by introducing a 
spatial coordinate $0< \s\leq 2\pi$, and
 $n(\s_a)= n({ 2\pi a \ov J})$, so that
 $n_{a+1} - n_{a} =  { 2\pi\ov  J} \d_\s n + \ldots$, etc.
Then we find (using Taylor expansion and
dropping a total derivative over $\s$)
\be \la{lety}
{ \l \ov (4 \pi)^2} \bra{n}\D_2\ket{n} \  \
 \to  \ \  { \bl \ov 8}   \bigg[ (\del_1 n)^2 + O({1 \ov J^2} \d_1^4 n)
 \bigg] \ , \ \ \ \ \ \ \ \ \
\  \bl = {\l\ov J^2} \ , \ \ \  \d_1 \equiv \d_\s \ ,  \ee
\be \la{wwty}
{ \l^2 \ov (4 \pi)^4}  \bra{n}\D_4\ket{n}\ \ \to \ \
 - { \bl^2 \ov 32}     \bigg[ (\del^2_1 n)^2 + O({1 \ov J^2} \d_1^6 n)
 \bigg]  \
 \ , \ee
 $$
{ \l^3 \ov (4 \pi)^6} \bra{n}\D_6\ket{n}
  \  \to \ \  { \bl^3 \ov   64}\bigg[\  { 7 \ov 4(2\pi)^2 } J^2
   (\del_1 n )^4
$$ \be\la{iqty}    +   \   (\del^3_1 n)^2
- { 19 \ov 24}  (\del_1 n)^2  (\del^2_1 n)^2
- \ { 115 \ov 12}  (\del_1 n \del^2_1 n)^2
 + O({ 1 \ov J^2 }\del^8_1 n )\bigg] \ .  \ee
 In general, starting with \rf{alta} one finds
 $$
{ \l^3 \ov (4 \pi)^6} \bra{n}\D_6(\a_1) \ket{n}
  \  \to \ \  { \bl^3 \ov   64}\bigg[ \ { 1 \ov 8(2\pi)^2}(14-\a_1)
   J^2 (\del_1 n  )^4
$$ \be\la{eqty}    +   \   (\del^3_1 n)^2
- { 1 \ov 48}(38 - 7 \a_1)   (\del_1 n)^2  (\del^2_1 n)^2
- \ { 5 \ov 24}(46+\a_1)  (\del_1 n \del^2_1 n)^2
+ O({ 1 \ov J^2 }\del^8_1 n 
)\bigg] \ .  \ee
 Here we used the identity \rf{ide} (for some useful relations 
 see 
  Appendix B).

We see that if $\bl$ is fixed in the large $J$ limit we may ignore higher
derivative terms in one-loop \rf{lety} and two-loop \rf{wwty} terms.
It is crucial for the consistency of our limit
that the subleading $(\del_1 n)^2$ terms cancel out in the 
higher-order $ \bra{n}\D_{2r+2}\ket{n}$ terms.\foot{This property
of $\D_{2r+2}$ should be a consequence of
supersymmetry of underlying SYM theory that restricts the
structure of the dilatation operator.}
We see explicitly
that this  is indeed the case for   $ \bra{n}\D_{4}\ket{n}$
(as was first  noticed  in \ci{kru})   and  for $ \bra{n}\D_{6}\ket{n}$.

However, the presence of ``subleading'' 
 $(\del_1 n)^4$ term
in \rf{iqty} that blows up in the $J\to \infty$ limit
implies a problem with taking the continuum limit
 directly in   $ \bra{n}D_6\ket{n}$.
Our conjecture is  that this singular term in \rf{iqty}  will
 be canceled out once quantum corrections are properly included.
 As we shall see in the next section, 
 one needs indeed to include order $n^4$ 
  quantum corrections, i.e. to   compute first an
effective  spin chain action and
 only then take  the continuum limit. 

  Assuming that the  ``scaling-violating''  $(\del n)^4$ term 
  in \rf{iqty}   can indeed be omitted, 
   we then finish with the following
 generalization of the semiclassical spin chain action \rf{gh}
\be \la{gth}
 \rS=  J  \int dt \inti\  \bigg[
   C_t(n)  - \langle\rH\rangle     \bigg]
       \ ,\ee 
       $$
 \langle\rH\rangle=  {\bl \ov 8}   ( \del_1 n)^2  
   - { \bl^2 \ov 32}  (\del^2_1 n)^2
 $$
 \be\la{kpu}
 + \ { \bl^3 \ov   64}\bigg[    (\del^3_1 n)^2
 - { 19 \ov 24}  (\del_1 n)^2  (\del^2_1 n)^2
-  { 115 \ov 12}  (\del_1 n \del^2_1 n)^2
\bigg] + O(\bl^4
)\ .   \ee
where we  used  the notation $\langle\rH\rangle$  instead 
of $\bra{n}\rH\ket{n}$
to indicate that some  quantum correction 
 was taken into account.

To 
compare  \rf{gth}  to  a similar  action \rf{kuru},\rf{kur},\rf{poj}
 obtained from the string \sm we need to rescale
  the time $t \to \bl^{-1} t$ as in \rf{resca} 
to absorb one (overall) power of $\bl= {1\ov \J^2}$. We then 
find perfect agreement of
 coefficients of  all
 quadratic  $(\del^r_1 n)^2$ ($r=1,2,3)$ terms.\foot{The agreement at
  quadratic order in $n$ 
 is related to having  the correct  BMN limit (see section \ref{sec:qc}). }
   However, there
 is no detailed agreement at $n^4$ and $n^6$ level. 
 In particular,
  \rf{kpu} is missing $\bl (\del_1 n)^4$ term present on the string side
  in \rf{kur}.
  The coefficients of the quartic  6-derivative terms
  in \rf{kpu}  are different from the ones
   in \rf{poj}; also, \rf{kpu} does not contain 
      $(\d_1 n)^6$ term.
   As we shall demonstrate below, quantum corrections
   on the spin chain side at order $\bl^2$
   induce  the term  $(\del_1 n)^4$  with precisely the same 
   relative coefficient
   $-{ 3\ov 4}$ as appearing  on the string side in \rf{kur}.
   
   Similarly, we expect that a systematic  account of
   quantum corrections  will be necessary to verify the 
   agreement of the  $n^4$ and $n^6$ 
   terms at the next   6-derivative order.
The inclusion of quantum corrections may turn out to be effectively
equivalent  to modifying the original dilatation operator 
by apparently non-local terms like $D_2^2,\ D_2 D_4,\ D_2^2,\ldots$.
As we shall show in Appendix C, their coherent state expectation values 
contain local parts very similar to the ones in \rf{wty},\rf{woty}.

Let us  note also that 
  the procedures of
  including quantum corrections and taking  continuum limit
   may not commute already at the first subleading order. 
   One may wonder that  if one first takes the continuum
   limit, then
   one gets the  $J$-factor in front of the action \rf{gth}
   and thus all quantum corrections
    would then be expected to  be suppressed  in the limit of
   $J\to \infty$. This, however, ignores a possibility of potential UV
   divergences  that are regularized away  in the discrete 
    spin chain
   but may  appear in the  continuum limit. 
   Since the  effective short-distance cutoff here 
   is essentially  $1\ov J$,  there may 
   be additional finite contributions
    coming from  divergent quantum corrections (due to cancellation
    of the $1\ov J$ suppression factor against the divergent cutoff 
    factor).
 Instead of trying to sort out such contributions
 in a continuum version, here we shall consider 
 directly the quantum version of the
 discrete theory, 
 derive a spin chain analog of  the quantum effective action, 
 and then take the  continuum limit.

\comment{NEW}
As we shall show  in Appendix C, second and third 
powers of $D_2$ operator have coherent state expectation values 
which look similar to those of $D_4$ and $D_6$.
In particular, to quadratic order in $n$
in the integrands   one observes that 
\be
\bran D_4 \ketn  = - \ha \bran  (D_2)^2 \ketn + O(n^4) \ , \ \ \ \ \ \ \ 
\bran  D_6
\ketn  = \ha \bran  (D_2)^3
\ketn  + O(n^4) \ .
\ee
We can make a conjecture similar to the one in 
\ci{bks}\foot{The conjecture in \ci{bks} 
was that, restricted to two-impurity BMN states, 
the exact dilatation operator has the form 
$D= J + 2 \sqrt{ 1 + { \bl \ov (4 \pi)^2} D_2} $
.}
and assume that 
for small fluctuations 
above the ferromagnetic 
ground state the exact dilatation operator  has the form 
(cf. \rf{dod}) 
\be 
\la{quadd}
D= J + \left[
\sqrt{ \idop + 2 {\l \ov (4 \pi)^2} D_2 } -\idop\right]
= J +  {\l \ov (4 \pi)^2} D_2
- \ha ( {\l \ov (4 \pi)^2} D_2)^2  + \ha  ({\l \ov (4 \pi)^2} D_2)^3 
+ ...   \ ,  
\ee
where $\idop$ indicates the identity operator.
Then, we   take the coherent state expectation  value 
of \rf{quadd},  keeping only the local terms (which are also the ones quadratic in $n$).
In the continuum limit this becomes:  
\be 
\bran D \ketn = J \inti  
\bigg[ 1 +  { 1 \ov 8} \bl (\del_1 n)^2 - { 1 \ov 32} \bl^2 
(\del^2_1 n)^2 +  { 1 \ov 64}\bl^3  (\del^3_1 n)^2  + ... \,\bigg]\ . \ee  
This 
is in precise  agreement with    the classical string 
expression \rf{quad}
that sums up all terms in the effective action that
are quadratic in $n$.


\setcounter{equation}{0}
\setcounter{footnote}{0}

\section{Energy of spin chain at order $\l^2$:\\
 quantum corrections 
to effective action } \label{sec:qc}


 In this section we compute the energy of the spin chain at order
$\lambda^2$ including quantum corrections. This gives the correct 
conformal dimension of the corresponding operators at the same order 
and therefore we now expect to reproduce what we obtained before from 
the string calculation, namely (\ref{kuru}),(\ref{kur}).

 In the same spirit of \cite{kru}, the states we are interested in
are represented  by spin waves with wavelength of order the size of
the chain $J$. These waves are going to be described by classical
solutions of a low energy effective action. Such 
a (``Wilsonian'')  action can, and
will, get contributions from integrating out the large momentum
modes. The calculation is therefore  more complicated than
at the leading 1-loop order  \cite{kru} 
 but the end result is that, after including the 
  quantum corrections
 and then taking the continuum limit, we reproduce 
 the string theory result (\ref{exep}),\rf{kur}.

\subsection{Spin chain Hamiltonian and first excited states}

Our starting point  is the Hamiltonian of the spin chain
proportional to the dilatation operator 
at order $\lambda^2$  \cite{bks}:
\be
\rH = {\bl}^{-1} \left(D_2 + D_4 \right) = - J^2  \left( \lambda_1 
\sum_{a=1}^{J} \left[ S^j_{a} S^j_{a+1}-\frac{1}{4}\right] 
 +  \lambda_2 \sum_{a=1}^{J} \left[
S^j_{a} S^j_{a+2}-\frac{1}{4}\right] \right)
 \ ,  \label{Ha}
\ee
where 
\beq
\lambda_1 \equiv  \frac{1}{4\pi^2} - \frac{1}{16\pi^4}
\lambda\ , \ \ \ \ \ \ \ \ 
\ \ \lambda_2 \equiv  \frac{1}{64\pi^4} \lambda\ .
\label{cconst}
\eeq
This is 
 the same expression as in  (\ref{dod}) and (\ref{iy}) after
 one uses  (\ref{Qdef}). We included  the factor
$\bl=\l/J^2$ between $\rH$ and the dilatation operator
$D=D_2+D_4$  to account for the fact that we shall 
 be assuming that time  $t$ is rescaled 
by $\bl^{-1}$   as in  \rf{resca}. 
As a result,  we get  the factors $J^2$ 
in front of $\rH$ which seem unconventional, but,
as we will see below, in the end make the 
perturbative expansion in $\bl$ more transparent.

 The Hamiltonian \rf{Ha} describes 
  a 1-dimensional spin $\half$ 
  ferromagnetic\foot{Note that the Hamiltonian is positive 
  for $\l_1 >0,\ \l_2 > 0$  (which is the case  assuming 
   $\l < 1$).}
   chain with first and second neighbor
interactions. The ferromagnetic ground state is
\beq
\ket{0} = \prod_{a=1}^J \ket{\uparrow}_a
\label{vacuum}
\eeq
with all spins parallel and therefore 
the total third spin projection 
is  $S_z = \ha J$.  The total spin
is $S=\ha J$ and so there are actually
 $2S+1 = J+1$ degenerate
ground states. The energy of the 
ground state is zero  in agreement
with the fact that the ground state describes a protected 
(chiral primary) operator tr$\Phi^J$ 
whose conformal dimension has no corrections to  any order in
perturbation theory.

Before proceeding,  it is  useful to 
find the first excited states
of the Hamiltonian \rf{Ha}. These are
 spin waves where one spin is down and all the others up. We can find
 these eigenstates exactly:
they are just given by momentum eigenstates:
\beq
\ket{k} = \frac{1}{\sqrt{J}}\sum_{a=1}^{J} e^{ika} |\uparrow
\uparrow \ldots \downarrow_a \ldots \uparrow\uparrow\rangle
\label{|k>}
\eeq
where we denote with $|\uparrow \uparrow \ldots \downarrow_a \ldots
\uparrow\uparrow\rangle$ a configuration with all spins up
except at site $a$ where it is down. It is easy to see that this
is an eigenstate of the Hamiltonian with eigenvalue
\beq
\epsilon(k) = J^2\left[ \lambda_1 (1-\cos k ) + 
\lambda_2 (1-\cos 2k)\right]\ ,  
\label{e(k)}
\eeq
which is   positive for  $\lambda_{1},\l_2>0$.

The next excitations correspond to two spins down and can be found
as superpositions of two spin waves. For large $J$ we can use a
dilute gas approximation and write those eigenstates as
\beqa\la{kkkk} 
\ket{kk'} &=& \frac{\sqrt{2}}{J} \sum_{a'>a=1}
^{J} e^{ika+ik'a'} |\uparrow
\uparrow \ldots \downarrow_a \ldots
\uparrow\uparrow\ldots\downarrow_{a'}\ldots\uparrow
\rangle \ , \\
\rH \ket{kk'} &\simeq& [\epsilon(k)+\epsilon(k')] \ \ket{kk'}\ , 
\eeqa
where the error is of order $1/J$\footnote{Also, $\ket{kk'}$ is
normalized up to factors suppressed by $1/J$.}. 
This approximation is good as
long as we assume that  the  number of spins down  
is much smaller than $J$.

\subsection{Defining the effective action } \label{sec:effacdef}

If we include a large number of spins down, which corresponds to
taking $J_1$ and $J_2$ to be of the same order,
i.e. of the same order as $J=J_1+J_2$, 
the correct description is in terms of an 
{\it effective action}  for low energy
modes with momenta  $\sim 1/J$.  To compute  it 
 we should divide the spin fields $S_a$ into
``slow'' and ``fast'' parts,
 with the slow modes being described by a unit
vector $\vn$  such that, 
when we take the lattice spacing $2\pi/J$ 
to zero, the derivative 
 $\partial \vn$ remains finite ($\partial\vn\sim 1$). 
 Taking into
account the rotational invariance,  the effective action for $\vn$
up to four derivatives should have the same 
form as in  (\ref{resca}) (in this section $\d\equiv \d_\s=\d_1$):
\beq
\rS = \rS_{\rm WZ} - J \int dt \istp  \left\{\frac{1}{8}(\partial \vn)^2
+ \bl \left[a_1 (\partial^2 \vn)^2 + a_2 (\partial \vn)^4
\right] + O(\d^6 n)  \right\}\  . 
\label{effa}
\eeq
The first term is the Wess-Zumino term which, upon quantization
ensures that we have a spin $\half$ at each site (\ie\ only two
states). The overall factor of $J$ comes from the spin chain length
and $J^2$ factor in \rf{Ha} gets  absorbed into the 
definition of the quadratic derivative term 
(the coefficient of $(\partial \vn)^2$ was fixed in \ci{kru}).
 We anticipated 
that the  4-derivative  terms are  of order $\bl$ but  
this should be an outcome of the 
calculation, namely we should get the coefficients 
$a_1,a_2\sim 1$.

Actually, the coefficients of all the terms quadratic in $\vn$ 
(i.e. $(\partial \vn)^2, \ (\partial^2 \vn)^2$, etc.) 
can be fixed by expanding $\epsilon(k)$ 
in \rf{e(k)} for small $k$ and comparing with the energy we get
from the effective action for small oscillations around
$\vn=(0,0,1)$. This corresponds in operator language to the BMN
limit.
This  small
oscillation analysis  does not,   however, 
 allow one to fix the non-trivial 
coefficient $a_2$  so we prefer to do
a direct computation of the effective action \rf{effa}.


 The effective action can be computed in various equivalent ways.
One usual method is to use the path integral formalism to
integrate out  the high energy modes. This leads to a diagrammatic
expansion. However, this approach presumes that one can separate
the action into a free and an interacting part with the free part
being quadratic in the fields. In our case we can use
$S^+=S_x + i S_y $ and $S^{-}=S_x - i S_y$ as our fields and replace
$S_z$  
through the identity 
\beq
S_z = \half -S^- S^+\ , 
\label{Szrep}
\eeq
valid for spin $\half$ (as can be seen by 
using the relation of $\vec S$ to Pauli matrices or by 
acting on the up and down basis:
$\{\ket{\uparrow},\ket{\downarrow}\}$).
The problem is that we get quartic interactions in the Hamiltonian with
no
small coupling constant in front. Instead, what is usually done for this
system 
is to consider a spin chain with arbitrary spin $s$ at each site. 
Near the state of maximum $S_z$ projection, we can expand $S_z$ as
\beq
S_z = \left[ \sqrt{s(s+1)} -(S_x^2+S_y^2)\right]^{\half} \simeq
s\left[1-\frac{1}{4s^2}(S^+S^-+S^-S^+)+\ldots\right]\ . 
\eeq
Now we get vertices of all orders in the fields $S^{\pm}$ but 
they are
suppressed in the limit
$s\rightarrow\infty$. It follows then  that  in this  limit
 we can treat these
interactions perturbatively
which is the content of the Holstein-Primakoff expansion \ci{hp}.

 This is, however,  not possible here since
 we definitely have to assume  $s=\half$ for the spin chain Hamiltonian
 to 
 represent the 
 SYM dilatation operator. Furthermore,  such an 
approach actually  obscures the simple nature of the system which
is obvious from  (\ref{vacuum}) and (\ref{e(k)}) where we found 
the ground state and first excited states with no difficulty. 
Nevertheless, as we will see at  the
end of this  section,  the final result seems
to be compatible with a large $s$ limit result 
 which suggest that an effective 
parameter of expansion is $Js$. 

We will leave further investigation of this
 issue for the future  and here will 
concentrate on an alternative approach which is suggested by the fact 
that the
effective action is the minimal value of the 
energy of the  state with fixed expectation values for the fields.

 Due to the condition $\vec{S}^2={3\ov 4}$ there are only two independent
 fields at each site so we need to fix two conditions on the expectation
 values. 
 It is  natural
to introduce a unit vector at each site $\vec{n}_a$ and look
 for the lowest energy state such that the mean value of the spin
 at each site $a$ points in the direction of $\vec{n}_a$. 
 That is,
 we describe the low energy wave as oscillations of the direction in
 which the spins point.
 This gives an effective action for $\vec{n}_a$ which we can then
 minimize by solving the
 classical  equations of motion. The field $\vec{n}_a$ is going to
 be considered as slowly varying in time and therefore to be a static
 background for the fast modes.
 Formally, what we want to find is the lowest energy state
 $\ket{\psi(\vec{n}_a)}$ such that
\beq
\bra{\psi(\vec{n})} \vec{S}_a \ket{\psi(\vec{n})}_{\perp} =0\ ,\ \ \ \ \
\ \ 
\ \ a=1, \ldots,  J\ ,
\label{parallelcond}
\eeq
where $\perp$ indicate the component of the vector in the
direction perpendicular to $\vec{n}_a$.

It is not possible to find such a state exactly, 
 so we need to resort
to a perturbation theory. Since we are interested in the limit of
large $J$ (long chains) it is natural to use $1/J$ as a small
parameter with  a requirement  of keeping terms of order
$\bl={\lambda\ov J^2}$ since we are interested in the limit \rf{lim} when
 this
quantity remains finite. It is natural to consider the 
lattice spacing to be $2\pi/J$
 so that the length of the chain is fixed
in the limit. 
The field $\vn$ represent modes whose wave-length is fixed
with respect to the length of the chain (but  grows to infinity in units
of the 
lattice spacing). More precisely, 
 we are to  keep $\partial \vn\sim 1$ in the limit.
 This means that the vectors $\vn_a$, $\vn_{a+1}$, at neighboring sites
 are almost
parallel. Thus, if we consider a state where, at each site, the spin is
aligned 
(\ie\ has maximum projection) along $\vn$, its energy will not differ
much from that of the vacuum. 
We can actually estimate the energy to be of order $E \sim J \int
(\partial \vn)^2 \sim J$.
\footnote{We remind the reader that we rescaled 
the time by $\bl^{-1}$ so  without  rescaling this energy is actually 
of order $J \bl= { \l \ov J} $.}   
 Such a state, constructed out of coherent states at each site
 (\ref{chs}), is a candidate
state to be the one of smallest energy such that
$\langle\vec{S}\rangle\parallel \vn$ and
is actually the one that was used in \rf{ety} (here we denote it
$\ket{\psi}_0$ instead of $\ket{n}$): 
\beq
\ket{\psi}_0 = \prod_a \ket{\vn_a} \ . 
\label{coherentstate}
\eeq
 We can correct this  state using perturbation
theory in $\bl$ to obtain
\beq
\ket{\psi(\vn_a)} = \ket{\psi}_0 + \ket{\psi}_1 + \ldots\ . 
\eeq
In this paper we are going to compute only the first
correction. 
The effective action for $\vn$ is then given by
\beq
S
 = S_{\rm WZ} - \bra{\psi(\vn_a)} \rH
\ket{\psi(\vn_a)}\ . 
\eeq
As was  already mentioned, the WZ term provides 
the correct quantization for $\vn$
(just $(2s+1)$ states at each site where here $s=\half$)
 and therefore, its coefficient cannot be renormalized 
 when integrating out the higher
momentum modes.\foot{Alternatively, in the path integral approach a
topological argument also implies that the coefficient of $\rS_{\rm WZ}$
is quantized in half integer units and therefore cannot be
renormalized.}

\subsection{Computing  the effective action } \label{sec:effaccomp}

Thus, to  the lowest order in the perturbative expansion (\ie\ with 
$\ket{\psi}\rightarrow\ket{\psi}_0$) 
the non-trivial part of the action
is thus determined   by the coherent-state expectation value  
(cf. \rf{kh},\rf{keh}) 
\beq
_0\bra{\psi} \rH \ket{\psi}_0 =-  \frac{1}{4} J^2 \left[
 \lambda_1 \sum_{a=1}^{J}
( \vn_{a} \vn_{a+1}-1) + \lambda_2
\sum_{a=1}^{J} ( \vn_{a} \vn_{a+2}-1)  \right] \ . 
\label{order1}
\eeq
This  happens to be  simply  the result found  by 
 replacing $\vec{S}\rightarrow
\ha \vec{n}$ in \rf{Ha} (see \rf{ety},\rf{rty}).
It is this expression  that we have already discussed in the previous
section
and which matched only partially  
the string result: it did not contain the $(\d \vn)^4$ term
(cf. \rf{wwty}).

Now we should compute the correction induced by the high energy
modes and see if it  contains  the missing terms.
Since the
coherent states (\ref{chs}) are defined  as 
\be
\ket{\vn_a} = R(\vn_a) \ket{\uparrow} \ ,
\label{chs1}
\eeq
where $R(\vn)$ is the rotation operator in \rf{rot}
it turns out to be useful  to define new operators at each site by 
\beq
\tS^i_a = R(\vn_a) S^i_a R^{-1}(\vn_a) = R_{ji}(\vn_a) S^j_a
\label{tSdef}\ , 
\eeq
i.e.  $\tS^i_a$ are 
  rotated with respect to the original operators $S^i_a$
  with the rotation matrix depending on the ``background field'' $\vn$. 
\foot{This step may be viewed as an analog of a covariant 
background-quantum 
  field split in the \sm.}
  $\tS^i_a$   obey the same
commutation relations,  and the zero-order state reads 
\beq
\ket{\psi}_0 = \prod_a \ket{\tilde{\uparrow}}_a\ . 
\label{newcoherentstate}
\eeq
Hence  $\ket{\psi}_0$ (which is the  product of coherent states
(\ref{coherentstate}))
 now  looks similar to the vacuum 
state (\ref{vacuum}).
 The condition
(\ref{parallelcond}) now reads 
\beq
\bra{\psi(\vn)} \tS^+_a \ket{\psi(\vn)} = \bra{\psi(\vn)} \tS^-_a
\ket{\psi(\vn)} =0\ , \ \ \
\ \ \ \ \ \ a=1,\ldots, J\ ,
\label{newparallelcond}
\ee
which we should now impose order by order in the perturbative expansion
in a similar way as we impose the normalization condition:
$$\bra{\psi(\vn)}\psi(\vn)\rangle=1 \ . $$
Everything now looks  simple except for the Hamiltonian that when 
written in terms of $ \tS^j_a $  
reads
\beqa
\rH &=& \rH_0 + V\ ,  \\
\rH_0 &=& -J^2 \sum_{q=1}^2 \lambda_q \sum_{a=1}^{J} \left[\tS^j_{a} 
\tS^j_{a+q}-\frac{1}{4}\right]\ ,  \\
V &=& - J^2 \sum_{q=1}^2 \lambda_q \sum_{a=1}^{J} 
A_{ij}^{a,a+q} \tS^i_{a} \tS^j_{a+q}\ , 
\la{ve}
\eeqa
where we introduced the $3\times 3$ matrices
\beq
A_{ij}^{a,a+q} = R_{li}(\vn_a) R_{lj}(\vn_{a+q}) - \delta_{ij}\ .
\eeq
These  are small if the variations of $\vn_a$ from site to site 
 are small (i.e. $\frac{1}{J}\partial \vn$ are small)  
  as we are going to assume below.
  
 The crucial point is that while the 
 Hamiltonian now is  more complicated it has naturally
separated into  ``large''  
 and   ``small''    parts.
Moreover, the state $\ket{\psi}_0$ is an eigenstate of
$\rH_0$. 
It is then easy to see that minimizing $\langle \psi | \rH |
\psi\rangle$ with respect to arbitrary corrections
to the ``ground state''   gives the usual
perturbation theory result
 where $V$ is considered a perturbation
of $\rH_0$.

 Let us briefly summarize how that happens. We consider
a state
\beq
\ket{\psi} = (1+c_0) \ket{\psi}_0 + \sum_{\ket{p}\neq\ket{0}} c_p
\ket{p} + \ldots\ , \ 
\eeq
where    $\ket{\psi}_0 \equiv \ket{0}$ and 
  $\ket{p}$ are eigenstates of $\rH_0$ with energy $\epsilon_p$.
The mean value of the energy is, to  first order,
\beq
\bra{\psi}\rH\ket{\psi} = \epsilon_0 + (c_0+c_0^*) \epsilon_0  +
\bra{0} V \ket{0} + \sum_p c_p \bra{0}V\ket{p} + \sum_p c^*_p
\bra{p}V\ket{0} + \sum_p |c_p|^2 \epsilon_p + \ldots\ , 
\eeq
where in our case $\epsilon_0=0$. 
Expressing $c_0$ in terms of  $c_p$ by taking into account that
\beq
\bra{\psi}\psi\rangle = 1\ \ \  \Rightarrow  \ \ \ c_0+c_0^* =
-\sum_p |c_p|^2\ , 
\eeq
and minimizing with respect to $c_p^*$ we get that
\beq
\ket{\psi} = \ket{\psi}_0 - \sum_p
\frac{\bra{p}V\ket{0}}{\epsilon_p-\epsilon_0}\ket{p} -\half
\sum_p\frac{|\bra{p}V\ket{0}|^2}{(\epsilon_p-\epsilon_0)^2}
\ket{0} + \ldots\ . 
\label{state1}
\eeq
>From here we get for the energy
\beq\la{epa}
\ep \equiv \bra{\psi}\rH\ket{\psi} = \epsilon_0 + \bra{0}V\ket{0} - \sum_p
\frac{\bra{0}V\ket{p}\bra{p}V\ket{0}}{\epsilon_p-\epsilon_0}
+ \ldots\ . 
\eeq
which is the standard   perturbation theory expression. 
The only
difference is that in our case we should be careful to include in
the sum over $\ket{p}$ only states such that $\ket{\psi}$
satisfies the condition (\ref{newparallelcond}).

 Next, it  is important to notice 
that since $V$ in \rf{ve} is quadratic in the 
spin operators we only need to
consider states which differ from the ground state by just one or
two spin flips. Those are precisely the states that we already
discussed above in subsection 4.1, so that, up to corrections
 of order $1/J$ we have 
\beqa
\ket{\psi} &=& \ket{\psi}_0 + \ket{\psi}_1 +\ldots \ , \\
\ket{\psi}_1 &=& \sum_k \alpha_k \ket{k} + \sum_{k,k'}
\alpha_{kk'}\ket{kk'}\ .
\eeqa
Then, to 
first order in the coefficients $\alpha_k$, $\alpha_{kk'}$, 
we   can  write condition (\ref{newparallelcond}) as
\beq
\bra{\psi}\tS^+_a \ket{\psi} =
\frac{1}{\sqrt{J}} \sum_k e^{ika} \alpha_k + ... =0\  ,  \ \ \ \ \  \
a=1, \ldots,  J \ . 
\eeq
This implies that 
$\alpha_k=0$, \ie\ we only need to consider states with two spin flips.
Evaluating  the energy \rf{epa} we get (remembering that here 
$\epsilon_0=0$)
\beqa\nonumber
\ep &=&  - \frac{J^2}{4}  \sum_{q=1}^2 \lambda_q \sum_{a=1}^{J}
A_{33}^{a,a+q} \\ &&  -J^4\frac{2}{J^2} \frac{1}{16} 
\sum_{a,a'=1}^{J} \sum_{kk'}
\sum_{qq'} \lambda_q\lambda_{q'} A^{a,a+q}_{++}A^{a',a'+q'}_{--}
\frac{
\bra{0}\tS^+_a\tS^+_{a+q}\ket{kk'}
\bra{kk'}\tS^-_{a'}\tS^-_{a'+q'}\ket{0}}{\epsilon(k)+\epsilon(k')}
\nonumber\\
&=& - \frac{J^2}{4}  \sum_{q=1}^2 \lambda_q \sum_{a=1}^{J}
A_{33}^{a,a+q} 
\la{nka} \rule{0pt}{1.8\baselineskip} \\ && -\frac{J^2}{8} \sum_{a,a'=1}^{J}\sum_{kk'}
\sum_{qq'} \lambda_q\lambda_{q'} A^{a,a+q}_{--}A^{a'a'+q'}_{++}
e^{i(k+k')(a'-a)}\frac{e^{ik'(q'-q)}}{2\epsilon(k')+(\epsilon(k)-\epsilon(k'))}
\nonumber
\eeqa
where the factor $1/16$ came  from the fact that the indices $+$,
$-$ contract with  coefficient $\half$ 
and the factor $2/J^2$ -- from the
normalization of the states $\ket{kk'}$ in \rf{kkkk}.

\subsection{Taking the continuum limit} \label{sec:contlim}

Since the field $\vec{n}$
is slowly varying we can now take the continuum limit and expand the
mean energy \rf{nka}
or effective action in derivatives of 
$\vec{n}$.
 In   Appendix D we give 
 the expansion of $A^{a,a+q}_{ij}$ in powers of $q\partial
\vn$ (this makes sense  since $q=1,2$ and $\vn$ varies
little between the neighboring sites).

The first term with $A_{33}$ is the same as the naive coherent 
state expectation value (and is thus quadratic in $n$), while 
 the leading  contribution to
$A_{++} A_{--}$ is already quartic in derivatives. If we ignore
the quantity $\epsilon(k)-\epsilon(k')$ in the denominator of the
last term in \rf{nka}, 
 the sum over $k+k'$ gives a delta function
$J\delta(a-a')$. If  instead we expand it in powers of $(k+k')$,
non-zero powers will give rise to derivatives of the coefficients
$A$ and therefore to higher than fourth  order derivatives. In view
of that we can consider just intermediate states with $k=-k'$ and
ignore the contribution of the rest of the states.
 Using  the relations from 
Appendix D 
\beqa
A_{++} &=& \frac{1}{2} \tilde{A}_{++} + \ldots\ , \ \ \ \ \ \ \ \ \ 
|\tilde{A}_{++}|^2=({2\pi q\ov J} \partial \vn)^4 + .... \ , 
 \\
A_{33} &=& -\frac{1}{2} \tpJq^2 (\partial\vn)^2 + \frac{1}{24}
\tpJq^4 (\partial^2\vn)^2 +\ldots\ , 
\eeqa
 we get 
\beqa\la{hpi}
\epsilon = J \istp\  \bigg[ d_0  (\partial\vn)^2 +  d_1  
 (\partial^2\vn)^2 +   d_2  (\partial\vn)^4  + ...\bigg]
 \ , 
\eeqa
where 
\beqa \la{oef}
d_0 &=& \frac{\pi^2}{2}  \sum_{q=1}^2 \lambda_q q^2 \ , \ \ \ \ \ \ \ \ \ 
d_1 = \frac{\pi^4}{6J^2}\sum_{q=1}^2 \lambda_q q^4\ , \\
\la{dee}
d_2 &=&  -\frac{\pi^3 }{8}
\sum_{q,q'} \lambda_q\lambda_{q'} q^2q'^2 
 \int^{2\pi}_0 dk \frac{e^{ik(q'-q)}}{\epsilon(k')} \ .\rule{0pt}{1.5\baselineskip} 
\eeqa
Here we have  replaced the sums over
 $k=2\pi n/J$, $n=1,\ldots, J$ and over
$a=1,\ldots, J$ by the integrals:
\beq
\sum_k \ \ \longrightarrow \ \ J\int_0^{2\pi} \frac{dk}{2\pi}\ ,
\ \ \ \ \ \ \ \ \ \ \ 
\sum_a \ \ \longrightarrow \ \ J\istp \ . 
\eeq
As follows from \rf{cconst}, 
\be \la{ioef}
d_0 =  \frac{ 1}{8}   \ , \ \ \ \ \ \ \ \ \ \ 
d_1 = \frac{\pi^4}{6J^2} \left( { 1 \ov 4 \pi^2 } -  { 3 \ov 16 \pi^4 } \l \right) 
=  { \pi^2 \ov 24 J^2} - {\bl \ov 32}    \ .  \ee
Omitting the first (subleading at large $J$) term in $d_1$, 
we thus get the same coefficients as in \rf{lety},\rf{wwty}.

Using (\ref{e(k)}) and expanding in powers of $\lambda$ we get
for the third remaining coefficient (which, up to $\bl$ factor, 
 should be equal to  
 the $a_2$ coefficient in (\ref{effa})) 
\beqa
d_2 &\equiv&  \bl\, a_2 
 =  - \frac{\pi^3}{8J^2} \int_0^{2\pi} dk
\frac{\lambda_1^2+8\lambda_1\lambda_2\cos k
+16\lambda_2^2}{\lambda_1(1-\cos k)+\lambda_2(1-\cos 2k)}
\\
&=&- \frac{\pi}{32J^2}\int_0^{2\pi}  \frac{dk}{1-\cos k}
+\frac{3}{2^{8}\pi}\bl \int_0^{2\pi} dk  -
\frac{1}{2^{11}\pi^3}J^2\bl^2\int_0^{2\pi} dk (3\cos k+5) +
\ldots\ ,  \rule{0pt}{1.8\baselineskip}\nonumber
\eeqa
where we substituted  the values of
 $\l_1$ and $\l_2$ from (\ref{cconst}).

Now few comments are in order. 
The first integral
over $k$ here is divergent. This is related to the fact that at this
order the interaction $V$ mixes two ground states and therefore
produces an IR divergence. 
We could correct this using  
perturbation to a degenerate level,  but, in any case this 
 contribution is 
of order $1/J^2$ so it is subleading  in the limit $J\rightarrow\infty$. 
The second finite term 
gives  precisely the  same coefficient 
\beq
a_2 = \frac{1}{32} \times \frac{3}{4} \ , 
\la{at}
\eeq
as required for the agreement with the string result in \rf{kur}.

One may worry though  about 
  the third and higher terms in the above expression that seem 
to dominate in the large  $J $ limit. 
However, since we did not include  higher order (three-loop, etc.) 
terms in $\bl$  in the Hamiltonian  \rf{Ha} the above  computation 
does not properly account for 
 the terms of order $\bl^2$. 
All such singular terms in $d_2$ should cancel 
after we include all the
terms $D_n$'s ($n\ge 3$) in the dilatation operator \rf{dod}. 
In fact,  as we 
saw in the previous section \ref{sec:expe}, 
there is  a similar singular $(\d \vn)^4$ contribution \rf{iqty} 
 coming 
from the naive continuum limit of the 3-loop operator  $D_6$.
 However,  to carefully check the cancellation of such singular terms 
we should also include other quantum corrections coming from $D_6$. 
Assuming  the required cancellations occur 
to all orders  so that the coefficient of $(\d \vn)^4$
is simply given by \rf{at}, 
we end up with the following expression for the effective 
 action to the 4-derivative order 
\beq
\rS = \rS_{\rm WZ} -J \int dt \istp 
\left\{\frac{1}{8}(\partial\vn)^2-\frac{\bl}{32}
\left[(\partial^2\vn)^2-\frac{3}{4}(\partial\vn)^4\right] +
 O(\d^6 \vn) \right\}\ , 
\label{Ssc}
\eeq
which is 
in perfect agreement with  \rf{bbb},(\ref{kur}),(\ref{ekdd}) or
\rf{loli}.

As a final comment let us 
note that if we  would have    done the above 
calculation for a spin chain 
with an arbitrary spin $s$ representation for the $SU(2)$ 
generators 
the coefficient $a_1$ would have been multiplied 
by $s^2$ and the coefficient $a_2$ -- by $s$. This
suggests that the $a_2 (\d n)^4$ term  we computed 
can be interpreted as a $1/s$ correction
in  the large $s$ expansion. It will be interesting 
to pursue this issue further
since it might indicate that the same calculation
 can be done directly in a
(regularized) continuum sigma model  set up.

\setcounter{equation}{0}
\setcounter{footnote}{0}

\section{Folded string  solution: a check} \label{folded}

\renewcommand{\ps}{\partial_x}
\newcommand{\ao}{a_0}
\newcommand{\bo}{b_0}


As a  check of the action \rf{Ssc} 
we have  derived  both from string theory and the spin 
chain let us show that it indeed correctly 
 reproduces the second-order ($\lambda^2/J^3$) correction 
 in  the expansion of the classical energy for the 
 folded string solution \cite{ft4,bfst}.
 On the spin chain side the  same ``two-loop'' term was found 
 using  Bethe ansatz technique  in \ci{serb}.

At lowest order this  check was performed in \cite{kru} where it
was shown that 
the corresponding  solution of the LL equation
of motion \rf{ll} following from the first and second derivative terms in
\rf{Ssc}  was given by
\beq \la{ansa} \phi=\omega t\ , \ \ \ \ \ \ \ \ \   \ \ \ \ \ \
\theta = \theta(\sigma)\ , 
\eeq
with
\beq
\ps\theta = \sqrt{\ao+\bo\cos\theta}\ , 
\label{tsol}
\eeq
where  $\theta$ and $\phi$ are polar coordinates for
the unit vector $\vn$ (see \rf{rot}).
The integration constants $\ao$ and $\bo$ should be expressed in terms
of the angular momenta $J=J_1+J_2$ and $\rS_3=(J_2-J_1)/2$, through 
\beqa
J&=&J_1+J_2 = 4\int_0^{\theta_0}
\frac{ d\theta }{\sqrt{\ao+\bo\cos\theta}} \\
\rS_3&=&\frac{J_1-J_2}{2} = -2\int_0^{\theta_0}
\frac{\cos\theta\ d\theta}{\sqrt{\ao+\bo\cos\theta}}\ .  
\eeqa
Here $\theta_0=\arccos(-\ao/\bo)$  (we assume $\bo>|\ao|$).
After that, the energy is computed by substituting  the values of 
$\ao$ and $\bo$  into
\beq 
E= \frac{\lambda}{8\pi^2}\int_0^{\theta_0}
\sqrt{\ao+\bo\cos\theta}\ d\theta\ . 
\eeq
All these
integrals can be done in terms of elliptic integrals and reproduce
the term of order $\lambda/J$ as obtained by  expanding the
energy of the exact folded rotating string  solution 
\ci{bfst}.

 To extend this to the next order we  shall start with 
  the action
(\ref{Ssc}).
Here it is useful to rescale back  the time 
$t$ and  to  reintroduce $\l=J^2\bl$. We also define  the
coordinate $x=\frac{J}{2\pi}\sigma$. The length of the chain is
now $J$. In this way we get an action 
\beq
 \rS = \rS_{\rm WZ} -  \int dt  \int_0^J dx
\left\{\frac{\lambda}{32\pi^2}(\partial\vn)^2-\frac{\lambda^2}{512\pi^4}
\left[(\partial^2\vn)^2-\frac{3}{4}(\partial\vn)^4\right]\right\}
\ , \label{Ssc2}
\eeq
 which leads to the modified Landau-Lifshitz equations:
\beqa
\partial_t \vn &=& -\ \frac{\lambda}{16\pi^2}
\vn\times\ps^2\vn+\frac{\lambda^2}{128\pi^4}\vn\times\ps^4\vn
\nonumber \\
&&+\ \frac{3\lambda^2}{256\pi^4}\left\{(\ps\vn)^2\,
\vn\times\ps^2\vn+2[(\ps\vn)\ps^2\vn]\,\vn\times\ps\vn\right\}\ . 
\eeqa
It is straightforward to check that the same ansatz \rf{ansa}
satisfies these  equations of motion provided 
\beq
\omega\sin\theta+\frac{\lambda}{16\pi^2}
\ps^2\theta-\frac{3\lambda^2}{256\pi^4}(\ps\theta)^2\ps^2\theta +
\frac{\lambda^2}{128\pi^4} \ps^4\theta=0\ . 
\eeq
We can simplify the equation by using $\theta$ as an independent
variable and introducing a new function $u(\theta)= (\ps
\theta)^2$. The resulting equation can be integrated once
resulting in
\beq
-256\pi^4\omega\cos\theta + 8\lambda\pi^2 u
-\frac{3}{4}\lambda^2u^2 +
\lambda^2  (uu''-\frac{1}{4}u'^2) = \tilde{a}
\eeq
where we used primes to indicate derivatives with respect to
$\theta$ and $\tilde{a}$ is an integration constant. In spite of being simpler this equation cannot be
integrated exactly. Doing a perturbative expansion in $\lambda$
(which is here equivalent to expansion in $\bl$) 
 we
get, to  lowest order
\beq
u = \frac{\tilde{a}_0}{8\pi^2} + 32\pi^2 \omega_0\cos\theta +
\frac{3\tilde{a}_0^2\lambda}{2048\pi^6} +
32\pi^2\lambda\omega_0^2+\frac{5\tilde{a}_0\lambda\omega_0}{4\pi^2}\cos\theta
+192\lambda\pi^2\omega_0^2\cos^2\theta+\ldots
\label{usol}
\eeq
Here we defined $\tilde{a}=\lambda \tilde{a}_0$ and
$\omega=\lambda\omega_0$ to reflect the fact that $\omega$ and $a$
are of order $\lambda$. 
Eq.(\ref{usol}) is required
only to know the relation between the parameters of the solution
and the angular velocity $\omega$. For all other purposes we can
simply write the solution, at this order, as
\beq
u(\theta) = (\ps\theta)^2 = a + b\cos\theta + c \cos^2\theta\ , 
\eeq
with
\beq
c = \frac{3\lambda}{16\pi^2} b^2\ . 
\eeq
The constants $a$ and $b$ follow from the condition that $J$ and
$\rS_3$ are fixed. We can compute them in an expansion $a=\ao+a_1
+\ldots$, $b=\bo+b_1\ldots$, $c=c_1+\ldots$ where $\ao$, $\bo$ are
the constants in (\ref{tsol}). The other constant $c$ is
already of order $\lambda^2$ since there is no $\cos^2\theta$ term
at lowest order.

The fact that we perturb the solution keeping $J$ and $\rS_3$ fixed
implies that
\beqa
\int_0^{\theta_1} \frac{d\theta}{\sqrt{a+b\cos\theta+c\cos^2\theta}}
 &=& \int_0^{\theta_0}
\frac{d\theta}{\sqrt{\ao+\bo\cos\theta}}\,\la{up}\\
\int_0^{\theta_1}
\frac{\cos\theta\ d\theta}{\sqrt{a+b\cos\theta+c\cos^2\theta}}\, &=&
\int_0^{\theta_0}
\frac{\cos\theta\ d\theta}{\sqrt{\ao+\bo\cos\theta}} \,\la{upp}
\eeqa
where, on the left hand side, $\theta_1$ is a zero of the denominator,
$a+b\cos\theta_1+c\cos^2\theta_1=0$. A straightforward but
lengthy calculation that
 we describe in Appendix E  shows that
this actually implies that at order $\lambda^2$ one also has 
\beq
\int_0^{\theta_1} \sqrt{a+b\cos\theta+c\cos^2\theta} \,d\theta \simeq
\int_0^{\theta_0}\sqrt{\ao+\bo\cos\theta}\,d\theta\ . 
\eeq
 This means that the energy evaluated from
the term with two derivatives (\ie\ $\sim (\ps\vn)^2$) is, at order $\l^2$, the same for 
the lowest order solution $(\ps\theta)^2 = \ao + \bo \cos\theta$ as it is for the
corrected solution $(\ps\theta)^2 = a + b \cos\theta+ c \cos^2\theta$. Therefore, 
the only non-vanishing $\lambda^2$ contribution comes from the evaluation of the term
quartic in derivatives on the  leading-order solution. This gives for
the corrected energy
\beq
\epsilon_2 = -\frac{\lambda}{512\pi^4}\int_0^{J} dx
\left[ (\ps^2\theta)^2 + \frac{1}{4} (\ps\theta)^4\right]\ , 
\eeq
which,  when evaluated on the unperturbed solution,  becomes
\beqa
\epsilon_2 &=& -\frac{\lambda}{512\pi^4}4 \int_0^{\theta_0} \frac{d\theta}{\ps\theta}
\left[ (\ps^2\theta)^2 + \frac{1}{4} (\ps\theta)^4\right] \mstrut{1.7}\\
&=&-\frac{\lambda^2}{512\pi^4}
\int_0^{\theta_0}\frac{\ao^2+\bo^2+2\ao\bo\cos\theta}{\sqrt{\ao+\bo\cos\theta}}
d\theta \mstrut{1.7}\\
&=& \frac{2}{\pi^4}
\frac{\lambda^2}{J^3}\rK_0^3\left\{(1-2x_0)\rE_0+(1-x_0)^2 \rK_0\right\} \mstrut{1.7}
\eeqa
where $\rE_0=\rE(x_0)$ and $\rK_0=\rK(x_0)$ are standard elliptic integrals and
\beq
x_0 = \frac{\ao+\bo}{2\bo}\ . 
\eeq
This is in perfect  agreement with the results obtained from
expanding the energy of the exact rotating string solution 
to  this
order \cite{ft4,bfst,serb}. 
\comment{NEW}
\foot{Similar result confirming the action \rf{Ssc}
by comparing to the energy 
of folded string 
solution  was 
independently obtained by A. Dymarsky and I. Klebanov 
(unpublished).}


\setcounter{equation}{0}
\setcounter{footnote}{0}

\section{Concluding  remarks}
 \label{remarks}

 In this paper we have shown that, up to order $\bl^2$,  
the anomalous dimensions of ``long''  two-scalar  operators in SYM 
theory can be obtained from a
semiclassical ``string'' 
action that precisely agrees with the expansion, to the same order,
of the string \sm  action in $R\times S^3 \subset AdS_5\times S^5$.
 Furthermore,  we have shown that,
if one is  able to compute the dilatation operator to  all loops,
 then one  can use a systematic
procedure to reconstruct a string action from  gauge theory. 
 
 Although this suggestion was already made in  \ci{kru} based on order $\bl$
 calculation and previous  results of \ci{bmn,ft2,bfst}, 
the general  procedure turns out to involve some  novel and non-trivial steps 
which are crucial to obtain agreement between gauge theory and string theory 
predictions   already at the first subleading  order $\bl^2$. 

The main idea is that
comparing the actions gives a map between configurations 
on both sides of the AdS/CFT duality
and therefore contains much more information than the comparison of 
energies of particular solutions. 
 Since one has to match  not only   the conserved charges (total energy and angular momenta) but
to actually map the variables on one side to the variables on the other side,
 gauge choices
and field redefinitions turn out to play a significant role. 
Our  gauge choice on the string side is motivated by the fact that the
angular momentum is uniformly distributed  along  the spin chain.
 Moreover, comparing 
other $SO(3)$  conserved charges also suggests the required  field 
redefinition which one can  make in a systematic
order by order way. 
On the spin chain side,  
we found that a naive semiclassical limit was not sufficient since  high energy modes
contribute to the low energy effective action starting at order $\bl^2$
(producing terms of quartic and higher order in the field).

 
Given the non-trivial agreement  between the spin chain and string  effective actions 
at the two leading orders 
 it seems natural to conjecture that, if one  could  sum the whole
perturbative series on the gauge theory side, the 
resulting effective action   would agree with the 
usual classical string action in $R\times S^3$ sector  of $AdS_5 \times S^5$. 


On the other hand,
the simplicity of the original ``unexpanded'' 
Polyakov's string  action in $R\times S^3$ 
seems to suggest that there may be  other  more efficient methods 
of  extracting  relevant ``string'' information from the gauge theory side.
In particular, 
let us mention that the idea of a low energy effective action 
for the Heisenberg-type  spin chain is much more 
powerful than what we actually used here. 
For example, 
using the BMN result on the gauge theory side  \ci{bmn,ZS}
 one can immediately fix
all (higher-derivative) terms {\it quadratic}  in $\vn$ in 
the  action. 
One  may  expect that the coefficient of the 
leading {\it quartic} 
term  $(\partial \vn)^4$   we have computed here
may be implicitly determining 
 the coefficients of all other higher derivative  terms  quartic in $\vn$.      

Leaving aside the possibility  of an all loop summation,
 there are many directions  one can 
investigate to put these ideas on firmer ground. 
 In particular, one would like 
 to  understand the role of subleading $1/J$ corrections. They 
should correspond to quantum $\a'$ corrections on the string side
(for which fermionic terms become important \ci{ft3}).
The fact that the
classical actions agree does not
a priori guarantee the agreement of these  subleading terms. 
\comment{NEW}
 It would also be interesting to understand the map between string
configurations and operators at higher orders. Let us recall how it
worked at lowest order \ci{kru}:
given a string configuration $n(\sigma)$ on the string side (in the 
coordinates of section 2) one constructs a spin chain $\prod_a \ket{n_a}$ with  
coherent states (by discretizing $n(\sigma)\rightarrow n_a$). Then one simply 
writes this state in the basis $\ket{\uparrow}$ and $\ket{\downarrow}$ of spins 
and identifies $\ket{\uparrow} \rightarrow X$, $\ket{\downarrow}\rightarrow Y$.
This gives a particular operator in the SYM theory that corresponds to the given 
string configuration. 
At the next order, that we analyze in the present paper, the map is already 
more involved since there is a field redefinition we need to do to put
the string action into the form required to match the spin chain effective
action. The field redefinition is suggestive of an identification between 
the mean value of the spin and the charge density associated to rotations
on the string side. Presumably, a better understanding of the map at higher loops 
would require making this identification precise.

\commentout{
{\bf I suggest to omit all below}

>From broader perspective, 
 study of sectors  with rotation on $AdS_5$ is also important. 
large $AdS_5$ spin $S$ where the conformal dimension grows logarithmically 
with $S$ \ci{gkp}\footnote{See also \cite{kru2} for an alternative
 derivation of this result.}, 
$\Delta=S+ f(\lambda) \ln S+\ldots$ is less understood since 
the coefficient is a non-trivial function of $\lambda$ which is only known in the
limits $\lambda\rightarrow 0$ and $\lambda\rightarrow\infty$. It would be interesting to 
see of the methods used here can shed light on that problem.
}

\commentout{
Quadratic fluctuation action: in \ci{art} we got
(a ``plane-wave'' type) action with constant coefficients expanding
near circular solution with constant radii.  May be it can be linked
to a specific limit of \sm action obtained from spin chain, i.e.
spectrum of fluctuations  will also  be captured and guaranteed to
agree this way.
}

\commentout{
Main issue: how to generalize spin chain to all loops-- here
\ci{kru} does not seem to give a clue. If we could do this, we could
hope to compute exact dimensions in $SL(2)$ sector of the full spin
chain and thus interpolating functions as coefficients
 $\ln S$ in large AdS spin $S$. It seems that so far all works
 only for several large spins -- that is relation to \sm;
 may be it is worth looking at continuous limit of $SL(2)$
 spin chain as well ?
} 

\section{Acknowledgments }
We are grateful to  G. Arutyunov, N. Beisert,
 S. Deser, 
A. Dymarsky, 
 S. Frolov, 
I. Klebanov, 
 J. Maldacena, A. Marshakov, J. Minahan, A. Parnachev,
J. Russo, 
  M. Staudacher
 and  K. Zarembo for useful 
discussions and/or  e-mail correspondence.
The work of M.K. was supported in part by NSF grants
PHY-0331516, PHY99-73935 and DOE grant DE-FG02-92ER40706
and that of A.R. in part by NSF grant PHY99-73935.
The  work of A.T. was supported  by DOE
grant DE-FG02-91ER40690 and the INTAS contract 03-51-6346.

 \newpage

 \def \do {{\d_1}}
\setcounter{equation}{0}
\setcounter{footnote}{0}
\setcounter{section}{0}

\appendix{Redefining away time derivatives}

Let us  explain the general procedure of elimination
of time derivatives from the action \rf{llii}
by local field redefinitions order by order in $1\ov \J$,
including the first two orders in expansion.
The key point  is that the variation of the leading term
in the action is proportional to $\d_0 n$.
Let us consider a  general  Lagrangian of field $n_i$
of the following structure
\be\la{str}
L= L_0 + \ep L_1 + \ep^2 L_2 + \ldots \ , \ \ \ \ \
L_1 = b_1 L'_0   + (L_1)_0 \ , \ \ \ \
L_2 = b_2 L'_0   + (L_2)_0 \ ,  \ee
where  terms with
$$L'_0\equiv  { \delta L_0 \ov \delta n}$$
are proportional to leading-order equations of motion
(in our case these are the terms containing  $\d_0 n_i$).
The notation $(\ldots)_0$ is used for  terms
obtained by applying  leading-order equations of motion
to eliminate $\d_0 n$ in terms of spatial derivatives.
$b_n$ may depend on $n$ and its derivatives.
$\ep$ is a small  expansion parameter ($1\ov \J$ in our case).
The idea is to make a field redefinition (we shall omit
tilda on redefined fields)
\be
n \to n+ \ep m_1 + \ep^2 m_2 + \ldots \ , \ee
so that to eliminate the above $ L'_0$-terms.
Expanding in $\ep$
$$
L=[ L_0 + \ep L'_0 m_1  + \ep^2 L'_0 m_2
+ \ha \ep^2 L''_0 m_1 m_1  + O(\ep^3) ] $$ \be\la{stra}
+ \ \ep \big[b_1  L'_0  + (L_1)_0  + \ep  b_1 L''_0 m_1 +
\ep  b'_1 m_1  L'_0  +    \ep ((L_1)_0)' m_1 + O(\ep^2) \big]
\ee
$$
+\  \ep^2 \big[ b_2 L'_0   + (L_2)_0 + O(\ep) \big] \ ,  $$
and requiring  $m_1 = - b_1$ we get
\be
L=  L_0   + \ep  (L_1)_0  +  \ep^2  \hat L_2 +
 O(\ep^3) \ ,   \ee
 \be\la{tre} \hat L_2 =
  L'_0 (b_2 +  m_2 - m'_1 m_1 )
 - \ha  \ep L''_0 m_1 m_1  + ((L_1)_0)' m_1 + (L_2)_0
 \ . \ee
The remaining $L'_0$ term can be canceled  by a proper choice
of $m_2$ term in the redefinition.
In the case we are interested in,  $L_1$ is quadratic
in time derivative, i.e. contains a term quadratic in $L'_0$.
Also, $L''_0$ may contain (up to integration by parts, and up to
extra spatial derivatives)  a term  proportional to
$L'_0$, i.e.
$$
m_1 =-b_1 =  c_1 L'_0  + (m_1)_0 \ , \ \ \ \ \ \ \ \
L''_0 = e_1 L'_0  +  (L''_0)_0 \ ,     $$
where $(\ldots)_0$ again means   part obtained by using the leading-order
equations of motion.
Then \rf{tre} becomes
$$ \hat L_2 =
  L'_0 \bigg[ b_2 +  m_2 - m'_1 m_1
  $$ $$ - \  \ep L''_0
  c_1 (m_1)_0  - \ha  \ep L''_0  L'_0 c_1 c_1
  + ((L_1)_0)' c_1    - \ha  e_1  (m_1)_0 (m_1)_0   \bigg]
$$
\be\la{trew}
 - \ \ha  (L''_0)_0  (m_1)_0 (m_1)_0
 + ((L_1)_0)' (m_1)_0 +
  (L_2)_0  \ . \ee
Choosing $b_2$ to cancel the first bracket
proportional to $L_0'$ we are left with
the following expression for the second correction
after the required field redefinition.
 \be\la{rew} \hat L_2 = (L_2)_0+ ((L_1)_0)' (m_1)_0
 - \ha   (L''_0)_0  (m_1)_0 (m_1)_0    \ . \ee
 It is thus not enough to drop
 terms in $L_2$ proportional to leading order equations of motion
 (i.e. to consider its $(L_2)_0$ part)
 and also to include a variation  of the first-order
 correction ($L'_1$ term): one also needs a term
 with second variation of $L_0$ (with the opposite sign to the
 one that would correspond to the expansion of $L_0$ term in
 $L$).
 In our case of \rf{llii}
 \be\la{lea}
 L_0 = C_0 - { 1 \ov 8  } (\d_1 n )^2 \ , \ \ \ \ \ \ \ \
 L_1 = {1 \ov 8 } (\del_0  n_i)^2   +  { 1 \ov 128 }
(\del_1 n_i)^4   \ , \ee
\be \la{tlii} L_2= -   {1 \ov   32} \big[
  (\del_0 n_i \del_1 n_i)^2 - {1 \ov 2}
 (\del_0 n_i)^2  (\del_1 n_k)^2
+ { 1 \ov 32 } (\del_1 n_i \del_1 n_i )^3  \big]
\ ,   \ee
\be \la{kj}
(L'_0)_i = \ha \ep_{ijk} n_j \del_0 n_k  +  \fo (\d^2_1 n_i)_\perp
  \ . \ee
To eliminate time derivatives to
first order  we need to do a field redefinition with
 \be \la{kp}
 m_1= - \ha\big[ \ha \ep_{ijk} n_j \del_0 n_k  -
 \fo (\d^2_1 n_i)_\perp \big]  \ , \ \ \ \ \ \ \ \ \
 (m_1)_0 = \fo (\d^2_1 n_i)_\perp \ .    \ee
 That gives $L_1 \to (L_1)_0$, i.e. the expression in \rf{kur}
 (again omitting tilda on redefined fields)
 \be\la{pov}
 (L_1)_0 = - H_1
  = {1 \ov 32 } \big( n''^2
  -   { 3\ov 4} n'^4     \big) \ , \ \ \ \ \ \ \
   \  n'_i \equiv \d_1 n_i \ .   \ee
   Below we shall need the relations following from $n^2=1$
   (prime on $n$   will stand for $\d_1$):
$n_i   n''_i = - n'_i n'_i$  and
\be\la{ttr}
 (nn'')'= -2 n''n'\ , \ \ \ \ \ \ nn'''= -3 n'n''\ , \ \ \ \ \
 (\ep_{ijk} n_i n'_j n''_k)^2 =
n'^2 n''^2  - (n'n'')^2 - n'^6 \ ,
\ee
 as well as
\be \la{ide}
 n'^2 (n'n''')
= - 2 (n'n'')^2 -   n'^2 (n'')^2  -  ( n'^2  n'n'' )' \ .
\ee
Using the leading-order equation of motion we have
\be \la{reqq}
(\del_0 n)^2 n'^2= \fo(  n''^2 - n'^4) n'^2\ , \ \ \ \ \
(n'\del_0 n )^2
= \fo [ n'^2 n''^2
- (n'n'')^2 -  n'^6 ] \ , \ee
so that
\be
(L_2)_0 =
 -  {1 \ov   64 } \big[{1 \ov 4} n'^2 n''^2 
- \ha (n' n'')^2  -  { 3 \ov 16 } n'^6 \big]  \ .
 \ee
We also find by direct computation (using  integration by
parts)
\be\la{hhh}
((L_1)_0)' (m_1)_0 \equiv
 {\delta (L_1)_0\ov \delta n}  (m_1)_0
= -{1 \ov   64 } \big[
n'''^2  - {5 \ov 2} n'^2 n''^2 - { 11}(n'n'')^2 + {3\ov 2} n'^6
\big]  \ . \ee
 Finally, we need to compute
$$
-\ha (L_0'')_0 (m_1)_0  (m_1)_0  \equiv
  - \ha   (  {\delta^2 L_0\ov\delta n \delta n} )_0  (m_1)_0 (m_1)_0
$$
\be =- { 1\ov 64}
\bigg[ \big( \ep_{ijk} (n''_i)_\perp n_j \d_0 (n''_k)_\perp\big)_0
+ \ha [ ((n''_i)_\perp)''(n''_i)_\perp
+ n'^2 ((n''_i)_\perp)^2 ] \bigg]
\ , \ee
where $(...)_0$ again means that
we are allowed to use the leading-order equations to
eliminate time derivatives in favor of spatial ones.
We then find, using  \rf{ide} and ignoring total derivatives
\be\la{joj}
-\ha (L_0'')_0 (m_1)_0 (m_1)_0=
-{1 \ov   64 } \big[
  {1 \ov 2} n'^2 n''^2 - (n'n'')^2 - {1\ov 2} n'^6
\big]  \ . \ee
Combining the above  three contributions together we end up  with
the following expression for $\hat L_2$ in \rf{rew}
\be\la{pojl}
\hat L_2 = - H_2=
-{1 \ov   64 } \big[ n'''^2
  - { 7\ov 4} n'^2 n''^2 - { 25
 \ov 2} (n'n'')^2
  + {13 
\ov 16} n'^6
\big]  \ . \ee

\setcounter{equation}{0}
\setcounter{footnote}{0}

\appendix{Some useful relations }
\def\ssl#1{\rlap{\hbox{$\mskip 3 mu /$}}#1}
\newcommand{\ft}[2]{{\textstyle\frac{#1}{#2}}}
\newcommand{\eqn}[1]{(\ref{#1})}

\newcommand{\bega}{\begin{array}}
\newcommand{\ea}{\end{array}}
\newcommand{\al}{{\alpha^\prime}}
\newcommand{\ale}{\alpha^\prime_{eff}}

\def \p {\partial}
\newcommand{\h}{{\hat h}}
\def \OO {{\cal O}}


To simplify the  expression for $\D_6$ in \rf{klp},\rf{alta} 
(i.e. to eliminate terms with repeated spins)
we  used the fact that spins at different sites commute,
and also  the  special relation between the spin 1/2
$SU(2)$  generators $S_i = \ha \s_i $
\beqa
\label{ins}
2 S_a^i S_a^j = [ S_a^i , S_a^j ] + \{ S_a^i , S_a^j \} = i
\ep^{ijk} S_a^k +
{1\over 2} \delta^{ij}\ . 
\eeqa
This  implies
\beqa
\label{ree}
(S_a \cdot S_b) (S_b \cdot S_c) &=& {1\over 4} S_a \cdot S_c +
{i \over 2}
\ep^{ijk} S_a^i S_c^j S_b^k
\ ,\\
\label{our}
(S_a \cdot S_b) (S_b \cdot S_c) (S_c \cdot S_d) &=&
{1\over 16} S_a \cdot S_d
+ {1 \over 4} (S_a \cdot S_d) (S_b \cdot S_c)
- {1 \over 4} (S_a \cdot S_c) (S_b \cdot S_d)
\nonumber\\&&
+ \ {i \over 8} \ep^{ijk} S_a^i S_d^j (S_b^k + S_c^k)\ ,
\eeqa
and similar relations.

In taking the continuum limit we noted that
$$
N_{a+p,a+q}\equiv  \ha ( n_{a+p} - n_{a+q})^2\ \ 
\to \ \ 
-\sum_{m=2}^\infty \sum_{l=0}^m {p^l q^{m-l} \over
l! (m-l)!} \; \p^l n \cdot \p^{m-l} n
$$ $$ 
= {(p-q)^2 \over 2} (\p n)^2
+ {(p+q) (p-q)^2 \over 2 } (\p n \cdot \p^2 n)
+  {(p-q)^2 (p+q)^2 \over 8} (\p^2 n)^2$$ $$ 
 + {(p-q)^2 (p^2 + pq + q^2) \over 6} (\p n \cdot \p^3 n) 
  +  O(\d^6 n ) $$ 
 \be \la{vevb}
= 
{(p-q)^2 \over 2! } (\p n)^2
- {(p-q)^4 \over 4! } (\p^2 n)^2
+ {(p-q)^6 \over 6! } (\p^3 n)^2 +  O(\d^8 n )
\ , 
\ee
where  $\d$ is derivative over $x= {J\ov 2\pi} \s$.
We omitted  total derivatives and used various identities
following from $n^2=1$.
>From this  we also find (again, integrating by parts) 
$$
N_{a+p, a+q}  N_{a+r, a+s} 
\ \ \to \ \ 
\fo {(p-q)^2 (r-s)^2  } (\p n)^4
$$
$$+\ \fo
{(p-q)^2 (r-s)^2 } (p+q)(r+s)
(\p n \cdot \p^2 n)^2  
$$ $$ + \ { 1\over 12 }(p-q)^2 (r-s)^2 [(p+q)^2 + (r+s)^2 - pq -rs ]
(\p n)^2 (\p n \cdot \p^3 n)  $$ 
\be
\label{red}+\ {1\over 16 }
{(p-q)^2 (r-s)^2 }[(p+q)^2 + (r+s)^2]
(\p n)^2 (\p^2 n)^2   + O(\d^8n)\ .
\ee

\setcounter{equation}{0}
\setcounter{footnote}{0}

\appendix{Continuum limit of coherent state\\  values 
of  product operators
}

Let us repeat the computations done 
 in section \ref{sec:expe} for the following
expectation  values of the products of $D_2$ and
$D_4$ operators in \rf{dod}.\foot{Although outside the main line of development of this paper,
we include these results partially motivated by analogous remarks in 
\ci{bks,serb}.} 
Let us define (subtracting disconnected parts of correlators)
\be \bar D_{2r} \equiv  D_{2r} - \bra{n} D_{2r} \ket{n}\ , \ee\
and consider 
 \be 
K_4\equiv \bra{n} \bar D^2  \ket{n}=
   \bra{n} (D_2)^2 \ket{n}  - (\bra{n} D_2 \ket{n})^2\ , \ee
$$
 K_6 \equiv  \ha \bra{n}( \bar D_2 \bar D_4 + \bar D_4 \bar D_2) \ket{n} 
$$
\be \la{kuy} =\ 
 \ha  \bra{n}  (D_2  D_4 +D_4  D_2)   \ket{n}   
 -  \bra{n} D_2  \ket{n} \bra{n} D_4  \ket{n}  \ ,
\ee  
$$
 \td  K_6 \equiv  \bra{n} (\bar D_2)^3 \ket{n} $$ 
 \be \la{hoj} =\  \bra{n} ( D_2)^3 \ket{n} - 3 
 \bra{n} (D_2)^2 \ket{n}  \bra{n} D_2 \ket{n} + 
 2 (\bra{n} D_2 \ket{n})^3 \ . \ee
One could expect a priori  that since each $D_{2r}$ factor
involves a sum over sites,   contributions of their products 
will be non-local in the continuum limit.
 However,  the above ``connected'' parts 
   turn out to contain 
local terms  which are very similar to the ones present in 
$\bra{n}D_4\ket{n} $ \rf{wty}  and $\bra{n}D_6\ket{n} $ \rf{iqty}.

Let us start with $K_4$, i.e. (cf. \rf{iy},\rf{Qdef})
\be\la{kkk}
K_4 =  
\sum^{J}_{a=1}
\sum^{J}_{b=1} \bra{n}(1 - \s_a\cdot \s_{a+1})  (1 - \s_b\cdot
\s_{b+1})\ket{n} -   [\sum^{J}_{b=1} \bra{n}(1 - \s_a\cdot \s_{a+1})\ket{n}]^2
 \ . \ee
 To evaluate this we split the sum over $b$ in the first term 
 into 3+1  terms:
 with $b=a, \ b=a+1, \ b=a-1$ (whose 
 sum will be denoted as $K^{(1)}_4$) 
 and the rest (denoted $ K^{(2)}_4$). The latter 
  is given by  (cf.  \rf{wty})
 $$
 K^{(2)}_4=  \sum^{J}_{a=1}
\sum^{J}_{b\not=a,a\pm 1}  N_{a,a+1} N_{a,a+1}  - (\sum^{J}_{a=1}
N_{a,a+1})^2  $$ 
\be\la{iir}
=  -
\sum^{J}_{a=1} N_{a,a+1}(N_{a,a+1} + N_{a-1,a} + N_{a+1,a+2})
=  - \sum^{J}_{a=1} N_{a,a+1}(N_{a,a+1} + 2N_{a-1,a}) 
  \ .  \ee
The sum of the first three terms  can be written as 
\be\la{hohh}
K^{(1)}_4 = \sum^{J}_{a=1}
\bra{n}(1 - \s_a\cdot \s_{a+1})
 \bigg[ (1 - \s_a\cdot \s_{a+1})  +  2 (1 - \s_{a-1}\cdot \s_{a}) 
  \bigg]\ket{n}\ ,  \ee
 where we  combined the terms with $b=a-1$ and $b=a+1$  by
 changing
 the summation variable from $a$ to $a+1$
 (in view of the  periodicity of the chain). 
 Simplifying this using \rf{ins},\rf{ree} we get 
 (note that $\ep^{ijk} n^i_a n^j_{a-1} n^k_{a+1}$ gives
   vanishing contribution)
   \be\la{hihh}K^{(1)}_4 = \sum^{J}_{a=1}
\bra{n} ( 6 - 8 \s_a\cdot \s_{a+1} + 2 \s_{a}\cdot \s_{a+2 }  ) 
 \ket{n}  = 2 \sum^{J}_{a=1} ( 4 N_{a,a+1} -  N_{a,a+2})
    \ . \ee 
    Comparing this to \rf{iy} we conclude that 
\be 
\bra{n} (D_2)^2 \ket{n} - (\bra{n} D_2 \ket{n})^2    = - 2  \bra{n} D_4 \ket{n} + K^{(2)}_4 \ .  \ee
The continuum limit is then found to be
 (including the $\l^2$ factor
and separating  the overall  power of $J$ with the integral 
over $\s$ as \rf{gth})
\be \la{conta}
 { \l^2\ov (4 \pi)^{4} } K^{(2)} \  \to\ \ 
 J \inti \ \bigg[  - { 3 \ov 64} \bl^2   (\del_1 n)^4 +  O( { 1 \ov
 J^2}\del^6_1 n ) \bigg] \ , \ee
i.e. (cf. \rf{wty}) 
\be \la{noni}
 {\l^2 \ov (4 \pi)^2} 
 \big[ \bra{n} (D_2)^2  \ket{n}  - (\bra{n} D_2 \ket{n})^2\big] 
  \  \to \ 
  J \inti \ \   {1 \ov 16} \bl^2 \big[
  (\d_1^2 n)^2  - { 3 \ov 4} (\del_1 n)^4 \big]  \ . \ee
It is a surprising coincidence that the local part of 
the coherent state expectation value 
of the operator $ - \ha  {\l^2 \ov (4 \pi)^2}   (\bar D_2)^2 $  
is thus the same
as required to match the string theory expression in 
\rf{kur}. Note also that as in $\bra{n} D_4 \ket{n}$
in \rf{wty}  here there is  no ``subleading'' 
$  (\del^2_1 n)^2$ term that would spoil the scaling limit.

 Similarly, we find 
$$
K_6
=\sum^J_{a=1} \big[
 -30 N_{a,a+1} + 12 N_{a,a+2} - 2 N_{a,a+3} $$
\be -\ (N_{a,a+2}- 4N_{a,a+1}) (  N_{a,a+1} +  N_{a+1,a+2} +
N_{a-1,a} ) - N_{a,a+2} N_{a+2,a+3}\big]\ , 
\ee
which has similar structure to (minus) 
$\bra{n} D_6 \ket{n}$ in \rf{woty}.
In the continuum limit then 
$$
{\l^3 \ov (4 \pi)^6}  K_6 \ \to \ \ 
 J \inti \  \  {\bl^3\ov   64 } 
  \bigg[ \  - { 1 \ov (2\pi)^2}  J^2  (\del_1 n)^4 
$$ \be\la{qoty} -  \  (\del^3_1 n)^2 +  { 7 \ov 6}  (\del_1 n)^2  (\del^2_1 n)^2   
+  { 25 \ov 3 }  (\del_1 n \del^2_1 n)^2 
 + O({ 1 \ov J^2} \del_1^8 n) \bigg] 
  \ .  \ee
 A much more involved computation gives 
$$
{\l^3 \ov (4 \pi)^6} \td K_6 \ \  \to \ \
 J \inti \ \  {\bl^3\ov  64  } \bigg[\  { 1 \ov (2\pi)^2}
   J^2 (\del_1 n)^4
 $$
 \be\la{koty}
  +\  2 (\del^3_1 n)^2  - { 31 \ov 6}  (\del_1 n)^2  (\del^2_1 n)^2
-  { 55 \ov 3 }  (\del_1 n \del^2_1 n)^2  
   +  {5 \ov 2} (\del_1 n)^6 + 
   O({ 1 \ov J^2} \del^8_1 n) \bigg]\ .   \ee
 The  linear combination 
  $ K'_6= 
  \ha (\bar D_4 \bar D_2 + \bar D_2 \bar D_4) +  (\bar D_2)^3 $ 
 thus has the coherent state expectation value 
 that  does  not contain the ``scaling-violating'' 
 $J^2 (\del_1 n)^4$ term. 
Note that  it  contains 
  $(\del_1 n)^6$ term (that was absent in  the expectation value of
  $D_6$ in   \rf{eqty}).
 
 The above discussion 
  may serve as an indication that a proper 
 account of quantum corrections 
 may produce an effective spin chain action 
 whose continuum limit will
  match the string theory result at the six-derivative order 
  \rf{poj}.

\setcounter{equation}{0}
\setcounter{footnote}{0}

\appendix{Rotation Matrices}

\newcommand{\bstrut}{\rule{0pt}{1.2\baselineskip}} 

Here we compute the matrices $A_{ij}$ that appear in section \ref{sec:qc}
and were useful
to calculate the quantum corrections to the action. By definition they 
are
\beq
A_{ij} = R_{li}(\vn_a)R_{lj}(\vn_{a+q})-\delta_{ij} =
\left(R^{-1}(\vn_a) R(\vn_{a+q})\right)_{ij} -\delta_{ij}\ , 
\eeq
where $R$ 
is the matrix of the rotation 
\beq
 R_{ij}(\vn_a) = e^{-i\phi_aS_z} e^{-i\theta_aS_y}
\eeq
in the vector representation. The angles $\phi_a$ and $\theta_a$
are polar coordinates of $\vn_a$ (see \rf{rot}).
The $ 3 \times 3$ 
matrix $A_{ij}$ can be evaluated as:
\beq
A = e^{i\theta S_y}e^{-i\Delta\phi S_z} e^{-i\theta'S_y}  
\eeq
resulting in
\beqa
A_{11} &=& \ct\ctq\cdp+\st\stq-1 \ , \ \ \ 
A_{12} = -\ct\sdp\ , \ \ \bstrut \nonumber\\ 
A_{13} &=& \ct\stq\cdp-\st\ctq\ , \bstrut \nonumber\\
A_{21} &=& \sdp\ctq      \ , \ \ \ 
A_{22} =\cdp-1 \ , \ \ \ A_{23} =\sdp\stq \ , \bstrut \\ 
A_{31} &=& \st\ctq\cdp-\ct\stq     \ , \ \ \ 
A_{32} = -\st\sdp \ , \bstrut \nonumber\\ 
A_{33} &=&\st\stq\cdp + \ct\ctq-1\ . \bstrut \nonumber
\eeqa
\newcommand{\sct}[1]{\scriptstyle{#1}}
\commentout{
\beq
A =  \left(\begin{array}{ccc}\sct{ \ct\,\ctq\,\cdp\,+\,\st\,\stq\,-\,1} & \sct{-\ct\,\sdp} &
\sct{ \ct\,\stq\,\cdp\,-\,\st\,\ctq} \bstrut\\ 
\sct{\sdp\,\ctq}  &  \sct{\cdp\,-\,1}  & \sct{\sdp\,\stq} \bstrut\\ 
\sct{\st\,\ctq\,\cdp\,-\,\ct\,\stq}  & \sct{-\st\,\sdp} & \sct{\st\,\stq\,\cdp \,+ \,\ct\,\ctq\,-\,1} \bstrut\\
\end{array}\right) 
\nonumber
\eeq}
where for brevity we defined $\theta=\theta_a$,
$\theta'=\theta_{a+q}$ and $\Delta\phi=\phi_{a+q}-\phi_a$. Since
the background field $\vn_a$ is slowly varying we can expand it in
derivatives. The components relevant for the calculations in section \ref{sec:qc} 
 are:
\beqa
A_{++} &=& A_{--}^* = (A_{11}-A_{22}) - i (A_{12}+A_{21}) 
  \simeq \half \tilde{A}_{++} + \ldots \\
A_{33} &=& \vn_a\vn_{a+q} - 1 \simeq -\half \tpJq^2 (\d_\s \vn)^2 +
\frac{1}{24} \tpJq^4 (\d_\s^2 \vn)^2 + \ldots
\eeqa
with
\beq
\tilde{A}_{++} = \tpJq^2 \left[-(\d_\s \theta)^2 + \sin^2\theta
(\d_\s \phi)^2 + 2 i \sin\theta\d_\s \theta\d_\s\phi\right]\ .
\eeq
In $A_{33}$ we omitted total derivatives since $A_{33}$ appears in the action 
integrated over $\sigma$.
Notice also that
\beq
|\tilde{A}_{++}|^2 = \tpJq^4 (\d_\s \vn)^4\ . 
\eeq

\setcounter{equation}{0}
\appendix{Integrals  for the folded string solution}

Here we want to prove the fact, used in section \ref{folded}, that the
energy at order $\lambda^2$ does not receive any corrections from
the term in the action which is quadratic in derivatives. This is
tantamount to showing
that eqs.\rf{up},\rf{upp} 
imply also  that at order $\lambda^2$
\beq
\int_0^{\theta_1} \sqrt{a+b\cos\theta+c\cos^2\theta}\  d\theta \simeq
\int_0^{\theta_0} \sqrt{\ao+\bo\cos\theta}\ d\theta \ . 
\eeq
 On both sides the upper limit of integration is a zero of the
corresponding function under the square root.
To check this fact we need to evaluate the following
integrals\footnote{These integrals
can be evaluated exactly in terms of standard elliptic functions $E(x)$
and $K(x)$ and then 
expanded for small $c$.} for small $c$:
\beqa
\lefteqn{ \int_0^{\theta_1}
\frac{d\theta}{\sqrt{a+b\cos\theta+c\cos^2\theta}}\, \simeq
\sqrt{\frac{2}{b}} K +
\frac{(b^2-2a^2+ab)K+2(2a^2-b^2)E}{\sqrt{2}b^{\frac{3}{2}}(b^2-a^2)}
c +\ldots}&& \nonumber \\
\lefteqn{ \int_0^{\theta_1}
\frac{\cos\theta\ d\theta}{\sqrt{a+b\cos\theta+c\cos^2\theta}} 
\simeq\sqrt{\frac{2}{b}} (2E-K) +} && \\
&& +\  \frac{\sqrt{2}}{3}
\frac{(b^2-4a^2)K+(8a^2+2ab-3b^2)E}{b^{\frac{5}{2}}(a+b)}c
-\frac{(b^2-2a^2+ab)K+2(2a^2-b^2)E}{\sqrt{2}b^{\frac{3}{2}}(b^2-a^2)}
c +\ldots  \nonumber \\
\lefteqn{ \int_0^{\theta_1}
{\sqrt{a+b\cos\theta+c\cos^2\theta}}\, d\theta 
\simeq\sqrt{\frac{2}{b}} \left(aK+b(2E-K)\right) +
\frac{(2a+b)K-4aE}{3\sqrt{2}b^{\frac{3}{2}}}c+\ldots} && \nonumber
\eeqa
where we defined 
\beq
K = K\left(\frac{a+b}{2b}\right), \ \ \ \ \ 
 \ \  E = E\left(\frac{a+b}{2b}\right),
\eeq
with $K(x)$ and $E(x)$ being standard elliptic integrals.  
In these expressions we should replace $a=a_0+a_1$, $b=b_0+b_1$ and
expand the result at first order in $a_1$, $b_1$. Demanding that
this first order terms in $J$ and $\rS_3$ cancel against the
contributions proportional to $c$ gives
\beqa
\frac{a_1}{c}&=&\frac{(\bo^2+\ao\bo-2\ao^2)K_0^2+(6\bo^2-8\ao^2)E_0^2+(8\ao^2-2\ao\bo^2-6\bo^2)K_0E_0}
{3\bo((\bo-\ao)K_0^2+2(\bo-\ao)K_0E_0+2\bo E_0^2)} \\
\frac{b_1}{c} &=& 2\frac{(\bo^2+\ao\bo-2\ao^2)K_0^2+2\ao\bo
E_0^2+(4\ao^2-2\bo^2-2\ao\bo)K_0E_0}
{3\bo((\bo-\ao)K_0^2+2(\bo-\ao)K_0E_0+2\bo E_0^2)}
\eeqa
where now
\beq
K_0 = K\left(\frac{\ao+\bo}{2\bo}\right), \ \ \ \  E_0 =
E\left(\frac{\ao+\bo}{2\bo}\right),
\eeq
Applying the same procedure to the energy and using the values of
$a_1$ and $b_1$ we have just obtained it is easy to show that the
first correction also cancels. This means that the term quadratic
in derivatives gives no correction to the energy at order
$\lambda^2$.


\end{document}